\documentclass[aps,prd,twocolumn,showpacs,floatfix,preprintnumbers,amsfont,amsmath,amssymb,nofootinbib, superscriptaddress]{revtex4-1}

\usepackage{color}
\usepackage[font=small]{caption,subfig}
\usepackage{hyperref}
\usepackage[normalem]{ulem}
\usepackage{lipsum}
\usepackage{url}
\usepackage{float}
\usepackage{ctable}
\usepackage{graphicx}
\usepackage[font=small,
   justification=justified,
   format=plain]{caption} 

\newcommand{\bt}[1]{\mathbf #1}

\newcommand{\sk}[1]{}

\newcommand{\refeq}[1]{Eq.~(\ref{eq:#1})}          
          
\newcommand{\reffig}[1]{Fig.~\ref{fig:#1}}          
\newcommand{\refsec}[1]{Sec.~\ref{sec:#1}}

\newcommand{\reftab}[1]{Tab.~\ref{tab:#1}}

\def\VEV#1{\left\langle #1 \right\rangle}

\newcommand{\be}{\begin{equation}}
\newcommand{\ee}{\end{equation}}
\newcommand{\ba}{\begin{eqnarray}}
\newcommand{\ea}{\end{eqnarray}}
\newcommand{\en}{\nonumber\\}

\newcommand{\apjl}{Astrophys. J. Lett.}
\newcommand{\aap}{Astron. Astrophys.}
\newcommand{\apjs}{Astrophys. J. Suppl. Ser.}

\newcommand{\mnras}{Mon. Not. R. Astron. Soc.}

\allowdisplaybreaks 

\begin{document}

\title{Effect of lensing magnification on the apparent distribution of  black hole mergers}

\author{Liang Dai}
\thanks{NASA Einstein Fellow}
\affiliation{\mbox{School of Natural Sciences, Institute for Advanced Study, 1 Einstein Drive, Princeton, New Jersey 08540, USA}}
\author{Tejaswi Venumadhav} 
\affiliation{\mbox{School of Natural Sciences, Institute for Advanced Study, 1 Einstein Drive, Princeton, New Jersey 08540, USA}}
\author{Kris Sigurdson}
\affiliation{\mbox{School of Natural Sciences, Institute for Advanced Study, 1 Einstein Drive, Princeton, New Jersey 08540, USA}}
\affiliation{\mbox{Department of Physics and Astronomy, University of British Columbia, Vancouver, British Columbia, V6T 1Z1, Canada}}

\date{\today}

\begin{abstract}

The recent detection of gravitational waves indicates that stellar-mass black hole binaries are likely to be a key population of sources for forthcoming observations. With future upgrades, ground-based detectors could detect merging black hole binaries out to cosmological distances. Gravitational wave bursts from high redshifts ($z \gtrsim 1$) can be strongly magnified by gravitational lensing due to intervening galaxies along the line of sight. In the absence of electromagnetic counterparts, the mergers' intrinsic mass scale and redshift are degenerate with the unknown magnification factor $\mu$. Hence, strongly magnified low-mass mergers from high redshifts appear as higher-mass mergers from lower redshifts. We assess the impact of this degeneracy on the mass-redshift distribution of observable events for generic models of binary black hole formation from normal stellar evolution, Pop III star remnants, or a primordial black hole population. We find that strong magnification ($\mu \gtrsim 3$) generally creates a heavy tail of apparently massive mergers in the event distribution from a given detector. For LIGO and its future upgrades, this tail may dominate the population of intrinsically massive, but unlensed mergers in binary black hole formation models involving normal stellar evolution or primordial black holes. Modeling the statistics of lensing magnification can help account for this magnification bias when testing astrophysical scenarios of black hole binary formation and evolution. 

\end{abstract}

\maketitle


\section{Introduction}
\label{sec:introduction}

The characteristic gravitational waves emitted during the inspiral, merger, and ringdown of a pair of black holes, long ago predicted by General Relativity, were recently detected at LIGO \cite{2016PhRvL.116f1102A, 2016arXiv160203840T}. This discovery conclusively demonstrates that stellar-mass black hole (BH) binaries exist, and hints towards a substantial population of binaries with masses as large as $\simeq 30\,M_\odot$ or beyond \cite{2041-8205-818-2-L22}.  In the near future  a new generation of ground-based detectors sensitive to lower strains (and possibly lower frequencies)  will come online \cite{2016arXiv160203838T,2015CQGra..32b4001A,2012CQGra..29l4007S,2010CQGra..27s4002P} and extend our gravitational wave (GW) reach to redshifts $z \gtrsim 1$ and unprecidently vast volumes of the Cosmos.  These prospects make such BH mergers one of the most exciting and important classes of events in all of astrophysics.

Binary mergers are standard sirens that reveal their luminosity distance on an event-by-event basis \cite{1986Natur.323..310S,Holz:2005df}. The standard mass-redshift degeneracy of GW astronomy is the statement that the frequency structure of gravitational waveforms cannot separate intrinsic mass-scales and redshifts. However, it is commonly assumed that a standard cosmological model can break this degeneracy and reveal both these quantities.\footnote{We can alternatively accomplish this via electromagnetic counterparts, but they are typically not expected to accompany stellar-mass black hole mergers.}

In this work, we emphasize that magnification due to gravitational lensing restores the mass-redshift degeneracy for BH merger events. The lensing magnification toward a given direction on the sky changes the luminosity distance as compared to the average cosmological value to the same redshift \cite{1992grle.book.....S}. Moreover, for high-redshift mergers, there is a non-negligible chance of strong lensing by intervening galaxies along the line of sight.

A magnified BH merger produces a physically identical response in a gravitational wave detector to an unlensed merger with a lower intrinsic redshift and larger intrinsic mass scale, but otherwise identical dimensionless parameters (e.g., mass ratio and spin). Thus, lensing biases the source redshifts and mass scales inferred using an assumed cosmology, even though it does not alter the waves' physical frequencies. We can trace this effect back to the geometric scale-free nature of General Relativity, which implies that the gravitational wave emission from black hole systems with different total masses but identical dimensionless parameters can be rescaled to a common form due to the absence of a mass scale in the spectrum. 

We {\it a priori} do not know whether an individual event has been strongly lensed or not. Even when multiple images are produced, due to the long time delays associated with galaxy lenses (of the order of weeks to months), they will trigger in the detector as separate events. The situation is different for mergers with identifiable EM counterparts, and hence redshifts, where only the inferred luminosity distance is confused. 

Several previous works have considered the astrophysical impact of GW lensing in other contexts. Lensing induces scatter in the Hubble diagram for standard candles or standard sirens~\cite{1993PhRvD..48.4738M,1997ApJ...475L..81W,2005ApJ...631..678H,Shang:2010ta,2010PhRvD..81l4046H}, while strong lensing in particular can leave a tell-tale high redshift tail in the neutron-star merger population as would be seen by LIGO~\cite{Wang:1996as} and the proposed Einstein Telescope (ET)~\cite{Piorkowska:2013eww}. Ref.~\cite{Biesiada:2014kwa} estimated the chances of obtaining a catalog of strongly lensed double-compact mergers using the ET, for several choices of binary masses. Lensing can potentially enable a statistical constraint of the Hubble parameter $H_0$, and facilitate the identification of host galaxies and extraction of source redshifts~\cite{Sereno:2011ty}. In the frequency band of space-based detectors the gravitational length scale of the lens can become comparable to the wavelength, and the resulting diffractive distortions of waveforms might enable better measurements of the lens parameters~\cite{Takahashi:2003ix}. To our knowledge, however, ours is the first study to expressly articulate the degeneracy caused by the combination of lensing magnification and the scale-free nature of gravity for BH mergers at cosmological distances.

We explore the statistical signatures of strongly magnified BH mergers with biased mass and redshift measurements. The relatively small optical depth to strong lensing suggests that such events are in general an insignificant population. In particular, the chances that GW150914, the first event detected, actually comes from a system of component mass significantly lower than $30\,M_\odot$ and from a redshift significantly higher than $z=0.9$ are tiny given plausible population models. However, strongly lensed events can generally  not be neglected if a large number of chirping bursts are  detected. We show that lensed events can alter the shape of the distribution of detected events if the intrinsic merger rate varies steeply with mass and redshift, an effect analogous to the strong-magnification induced tail in galaxy or quasar luminosity functions~\cite{1984ApJ...284....1T,Wyithe:2002ui,2011Natur.469..181W}. Both the increase in available comoving volume, and astrophysical effects like the evolution of the star-formation rate and metallicity, can generically skew the merger rates for BH binaries at the median mass of a source population at high redshifts. In particular, if the intrinsic merger rate has a strong physical cutoff in mass and/or redshift, lensed high-redshift events may in fact dominate over genuinely massive mergers at lower redshifts. We demonstrate the importance of this effect when testing astrophysical models of BH binary production and evolution against forthcoming observations.

We structure this paper as follows: In \refsec{degeneracy}, we demonstrate the mass-redshift-magnification degeneracy and quantify its size. In \refsec{mpdf}, we present a model for the probability distribution of the magnification due to galaxies embedded in the large-scale structure of our standard cosmology. In \refsec{statistics}, we derive lensing's effect on the distribution of observable events, and then in \refsec{len-highm-tail} we study this effect in the context of a number of plausible astrophysical models of the BH binary population and its evolution. For each scenario, we discuss the extent to which lensing magnification complicates the interpretation of population statistics. We conclude with a discussion of future directions in \refsec{discussion}, wherein we highlight the potential of multiple imaging to disentangle the effects of lensing from intrinsic variation of source populations, and thus explore the intrinsic merger history of BH binaries.

Throughout this paper we assume a fiducial flat $\Lambda$CDM cosmology with $\Omega_m=0.27$ and $h=0.7$. In all sensitivity calculations, we use GW waveforms that were computed according to the IMRPhenomC approximant \cite{2010PhRvD..82f4016S}. We use the definitions of the dimensionless characteristic strains and noise amplitudes outlined in Ref.~\cite{2015CQGra..32a5014M}, and we also employ `RMS characteristic strains' which are directly related to the root mean square (RMS) signal-to-noise averaged over all orbital inclinations and sky-locations \cite{1993PhRvD..47.2198F}.

\bigskip

\bigskip       
    

\section{Lensing and parameter degeneracy}
\label{sec:degeneracy}

We adopt the geometrical optics approximation for lensing, i.e., we assume that the propagating gravitational waves have wavelengths that are much shorter than the spatial length-scales associated with the intervening lens.\footnote{The relevant length scale for a lensing potential is the Schwarzschild length corresponding to the lens mass, not the lens' physical extent or Einstein radius~\cite{Takahashi:2003ix}.} In this limit, lensing has two effects: deflection of null geodesics by the lens' gravitational field, and change of the cross-sectional area of infinitesimal ray bundles (the latter effect rescales the energy flux by a magnification factor $\mu>0$, as in the case of electromagnetic waves). In Appendix \ref{app:waveeffects}, we demonstrate the validity of this approximation for the lensing of GWs from stellar-mass mergers by foreground galaxies or clusters. 

Let us assume that the background cosmology has a distance-redshift relation $d_L(z)$, where $d_L$ is the luminosity distance. Now consider a binary merger with intrinsic mass-scale $M$ that occurs at redshift $z$.\footnote{The mass scale $M$ may be any chosen combination of the component masses $M_1$ and $M_2$ (e.g., the chirp mass $(M_1 M_2)^{3/5}/(M_1+M_2)^{1/5}$).} Suppose the intervening mass distribution lenses the GWs with a magnification factor $\mu>0$. The strain amplitude is amplified by a factor of $\sqrt{\mu}$.

If we are ignorant of the lensing magnification, we can still fit the observed waveform to an inferred mass scale $\tilde M \neq M$ and an inferred source redshift $\tilde z \neq z$ (with all dimensionless parameters, e.g., the mass ratio and the spin parameter, unchanged). Since lensing does not affect frequencies, and thus the redshifted mass-scales of the waveforms, we have the mass-redshift degeneracy
\ba
\label{eq:Mzdegen}
\tilde M\,\left(1+\tilde z\right) & = & M\, \left(1+z\right).
\ea
The observed quantity is the characteristic strain $h_c(f_o)$ at every {\it observed} frequency $f_o$. It is given by~\cite{Salcido2016} 
\ba
h_c(f_o) & = & \sqrt{\mu}\,\sqrt{\frac{2G}{c^3}}\,\frac{1+z}{\pi\,d_L(z)}\,\left[\frac{dE}{df_s} \right]^{1/2}_{f_s=f_o(1+z),M},
\ea 
where $(dE/df_s)_{f_s,M}$ is the radiation energy spectrum for a source with an intrinsic mass scale $M$, expressed in terms of the {\it intrinsic} frequency $f_s=f_o(1+z)$. We then have the relation
\ba
\label{eq:ampdegen}
\frac{1+\tilde z}{d_L(\tilde z)} \left[\frac{dE}{df_s}\right]^{1/2}_{f_o(1+\tilde z),\tilde M} = \sqrt{\mu}\,\frac{1+z}{d_L(z)} \left[\frac{dE}{df_s} \right]^{1/2}_{f_o(1+z),M}.
\ea

\begin{figure}[t]
  \begin{center}
    \includegraphics[width=\columnwidth]{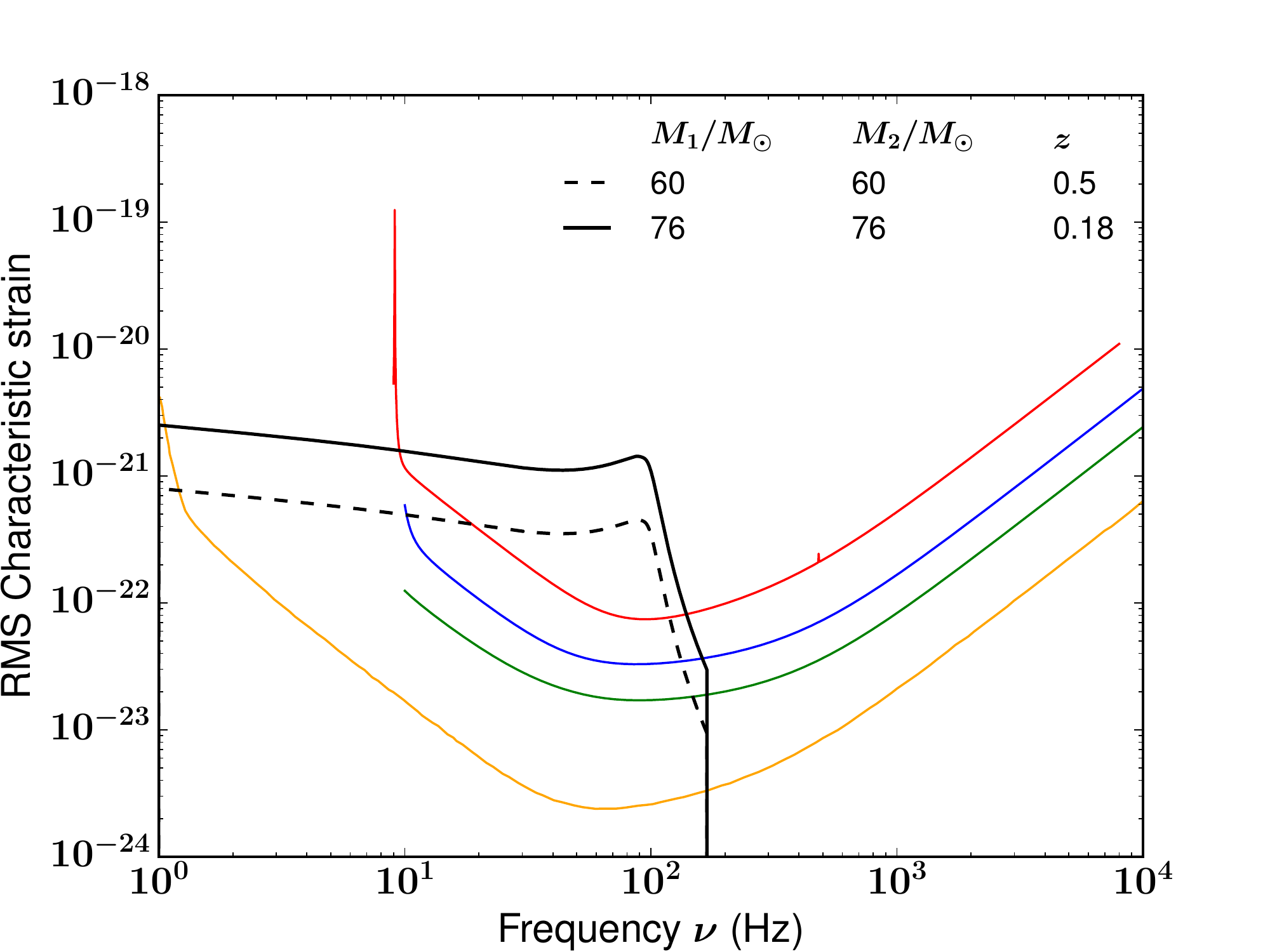}\\
    \vspace{0.1cm}\hspace{-0.15cm}\includegraphics[scale=0.56]{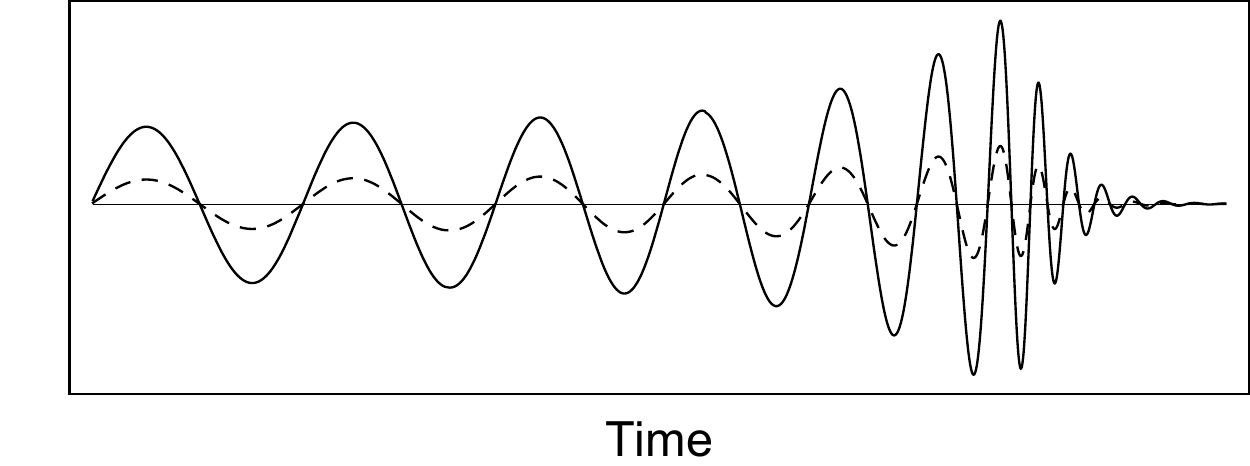}
    \caption{\label{fig:degeneracy}Illustration of the lensing-induced degeneracy: {\it Upper}: The solid and dashed thick black curves show the RMS characteristic strains for high- and low-mass mergers, respectively. These are perfectly degenerate if the latter is magnified by a factor $\mu = 10$. Also shown are the noise amplitudes for three stages of the LIGO detectors: {\tt current} (red), {\tt design} (blue), and {\tt ultimate} (green), and for the proposed Einstein Telescope {\tt einstein} (orange). {\it Lower}: Waveforms corresponding to the two chosen mergers. If the lower-mass merger (dashed) were magnified by a factor of $\mu=10$, the two waveforms would overlap.}
  \end{center}
\end{figure}

Since General Relativity is a geometrical theory, the vacuum Einstein equations are invariant under a rescaling of all masses, along with an accompanying rescaling of the spatial and temporal scales. This invariance guarantees that
\ba
\label{eq:muMzdegen}
\frac{1+\tilde z}{\tilde M} \left[\frac{dE}{df_s}\right]_{f_o(1+\tilde z),\tilde M} = \frac{1+z}{M} \left[\frac{dE}{df_s}\right]_{f_o(1+z),M}.
\ea
By substituting Eqs.~\eqref{eq:Mzdegen} and \eqref{eq:muMzdegen} into \refeq{ampdegen}, we observe that a magnification $\mu$ is equivalent to a rescaling of the luminosity distance, i.e.,
\ba
\label{eq:mudLdegen}
d_L(\tilde z) & = & d_L(z) /\sqrt{\mu}.
\ea
Equations \eqref{eq:Mzdegen} and \eqref{eq:mudLdegen} together characterize the observational degeneracy between lensed and unlensed mergers. Note that even though lensing does not {\it physically} alter the wave frequency, it affects our estimation of both the mass and redshift. This happens because we use the background cosmology for parameter estimation, while the presence of a lensing potential along the line-of-sight effectively alters the cosmology in that direction. 

This degeneracy is irresolvable for BH mergers without any independent redshift estimates. We can break this degeneracy for compact stellar mergers, such as those involving neutron stars, by applying theoretical priors on the masses or extracting redshifts from their EM counterparts.

\begin{figure}[t]
\centering
\hspace{-1cm}
\includegraphics[scale=0.46]{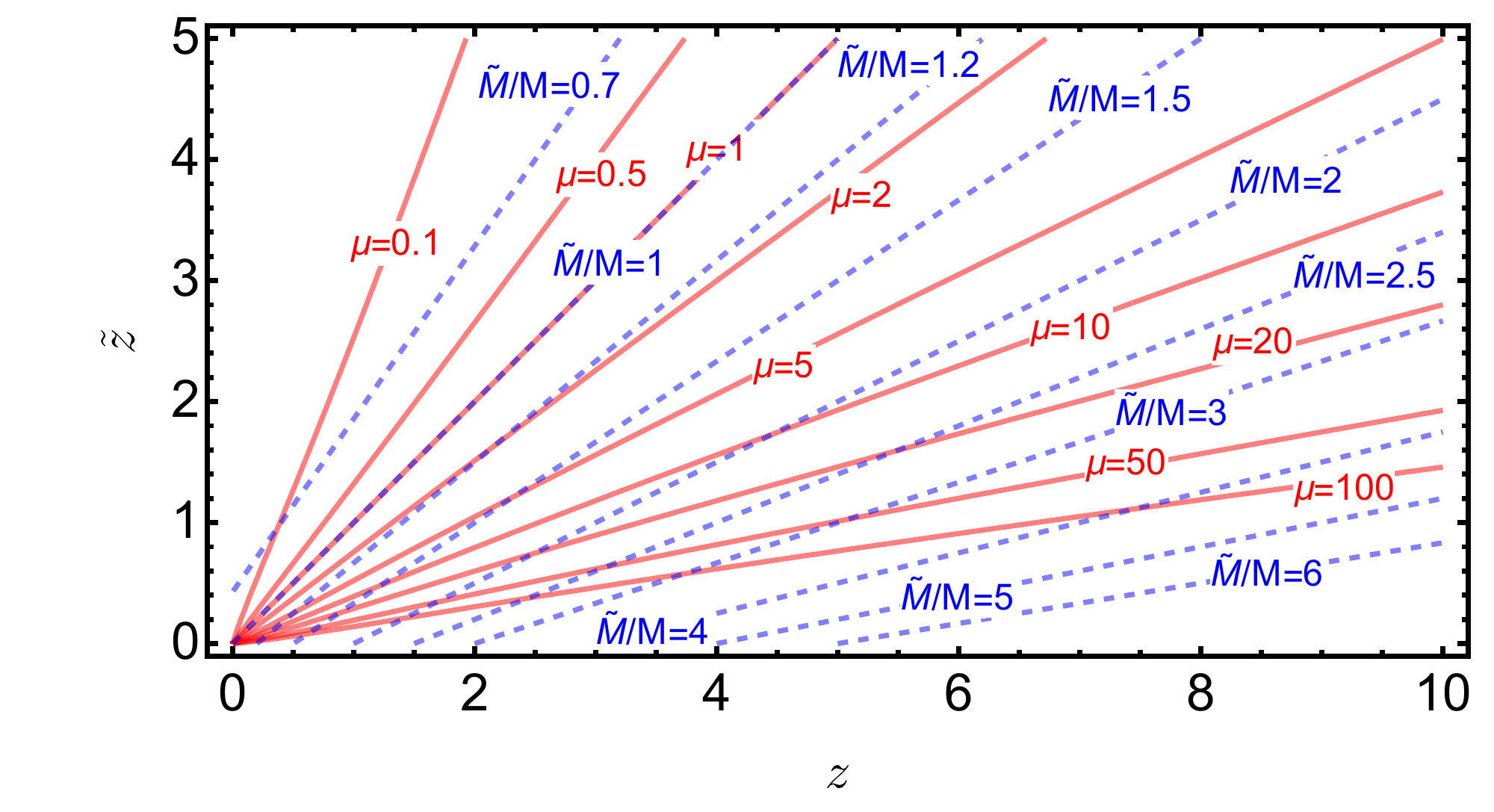}\\
\hspace{-1cm}
\includegraphics[scale=0.46]{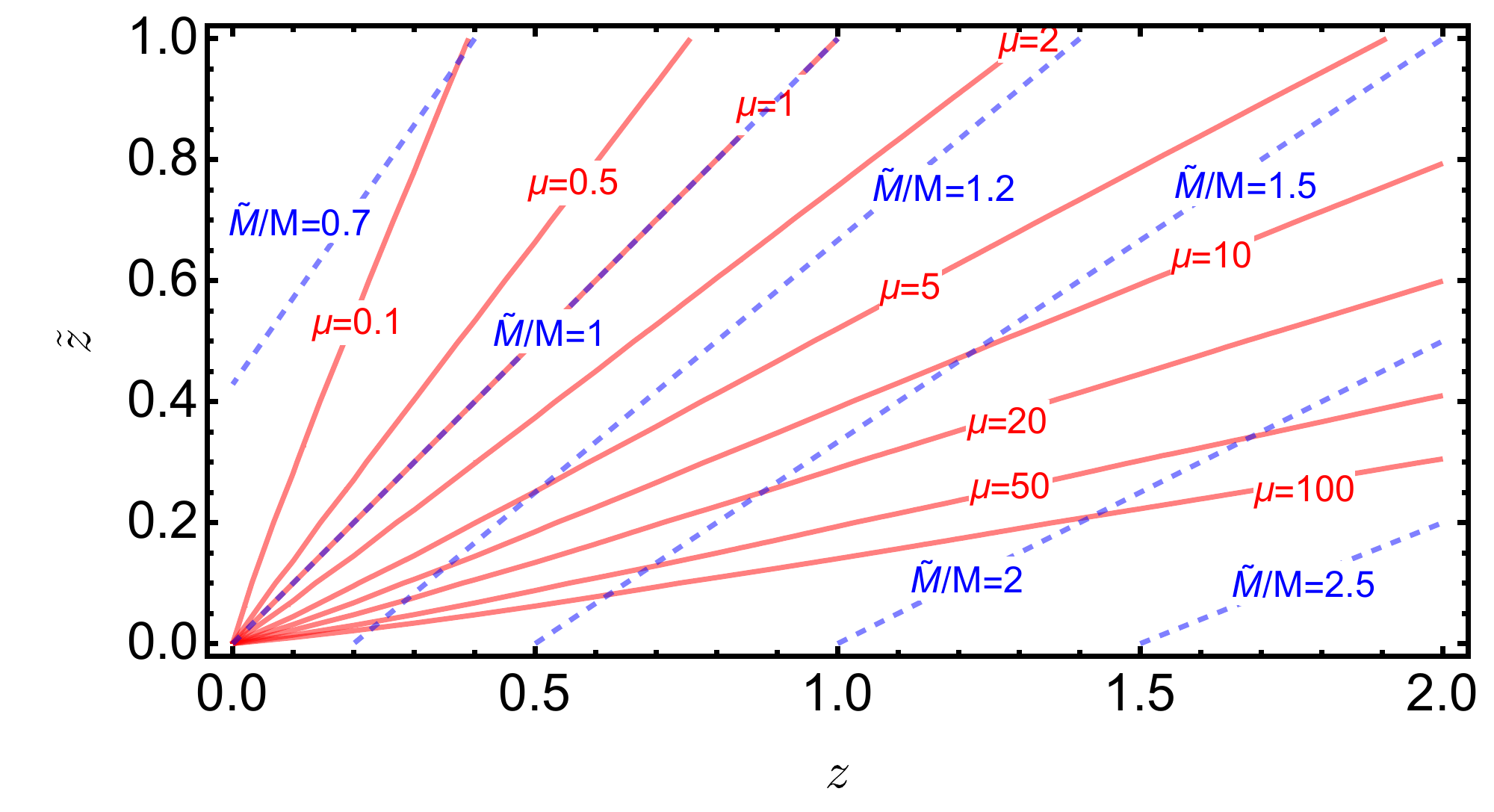}
\captionof{figure}[]{{\it Upper}: Mapping between the intrinsic redshift $z$ and inferred redshift $\tilde z$ for our fiducial cosmology. We show contours of constant magnification $\mu$ (red solid lines) and constant ratio of inferred and intrinsic masses $\tilde M/M$ (blue dashed lines). {\it Lower}: The same plot with scales chosen to emphasize low redshifts.}
\label{fig:ztotildez}
\end{figure}

Figure \ref{fig:degeneracy} shows the waveforms from an equal-mass BH binary merger with component masses $M_1 = M_2 = 60~M_\odot$ at redshift $z=0.5$, and another merger with masses $M_1 = M_2 = 76~M_\odot$ at redshift $z=0.18$. The characteristic strain waveforms are perfectly degenerate if the high-redshift merger has a magnification of $\mu=10$. The figure also shows noise amplitudes for four different detectors as benchmarks~: (1) current LIGO ({\tt current}) (2) LIGO at its design sensitivity ({\tt design}) (3) an optimal upgrade of LIGO using accessible technologies ({\tt ultimate}) \cite{Miller:2014kma}, and (4) the proposed Einstein Telescope (ET) ({\tt einstein}) at its design capability~\cite{Hild:2008ng}.

Figure \ref{fig:ztotildez} illustrates the mapping between intrinsic and inferred parameters as a function of the magnification. As one example, the gravitational wave event GW150914~\cite{2016PhRvL.116f1102A} could (in principle) be due to a BH pair with masses $\simeq 15 M_\odot$ instead of $\simeq 30 M_\odot$, if the merger occurred at redshift $z \simeq 1.2$ (instead of  $z \simeq 0.1$) and was magnified by $\mu \simeq 300$. For low-redshift mergers, the error in the estimated intrinsic masses is insignificant except for inconceivably large magnifications. The situation changes at high redshifts, where large but plausible magnification factors ($\mu \sim \mathcal{O}(10)$) can result in order-unity error in the estimated intrinsic masses. For instance, an equal-mass merger with component masses $\simeq 15 M_\odot$ that occurs at $z \simeq 4$ can appear as a massive merger (masses $\simeq 30 M_\odot$) from $z \simeq 1.3$, if it is magnified by a factor of $\mu \simeq 10$.

\begin{figure}[t]
\centering
\hspace{-0.5cm}
\includegraphics[scale=0.65]{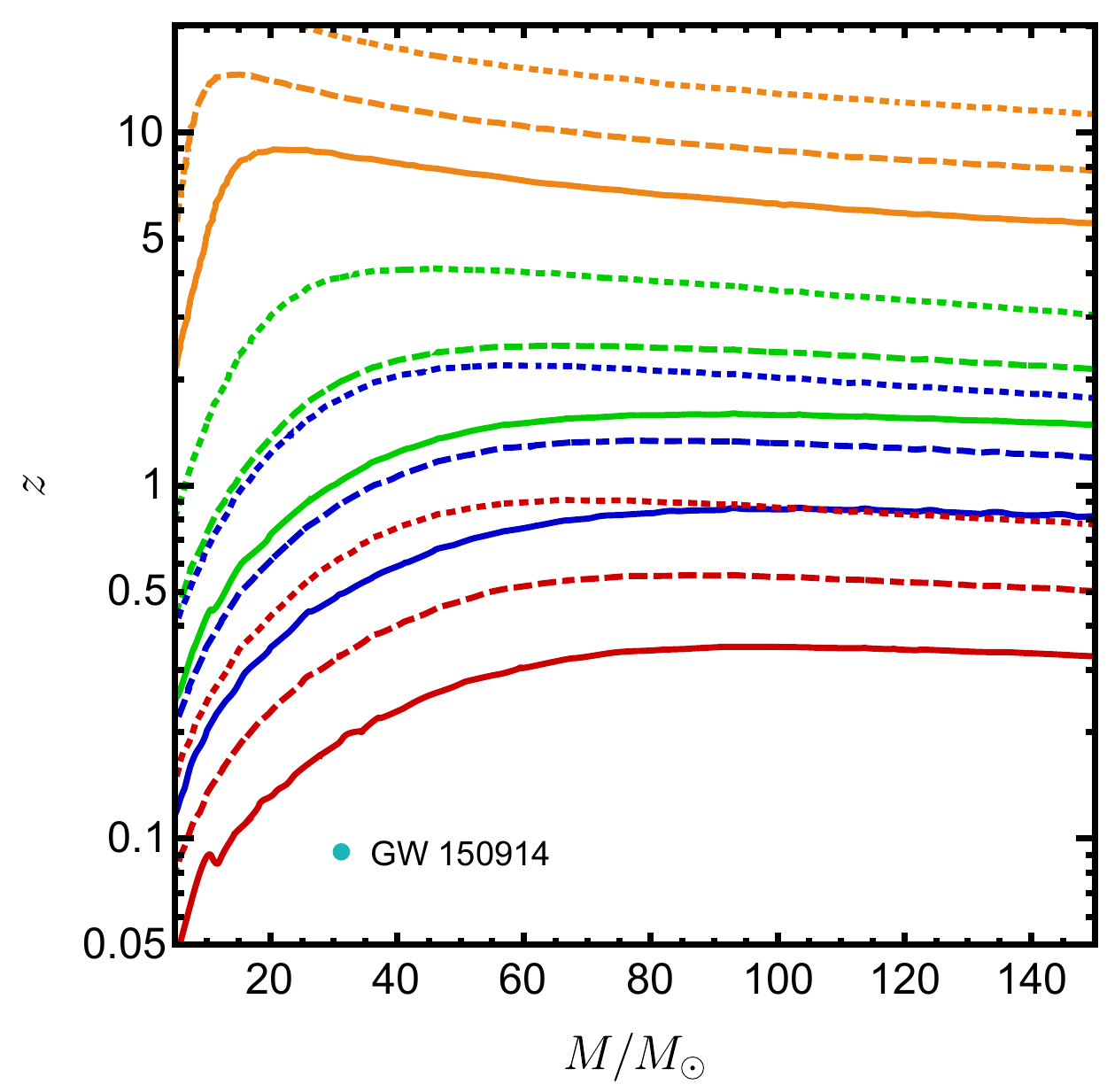}
\caption[]{Effect of magnification on the detection limit (SNR$=8$) for equal-mass BH mergers as a function of the intrinsic component mass $M$ and intrinsic source redshift $z$. For LIGO, we study three sensitivities: {\tt current} (red), {\tt design} (blue), and {\tt ultimate} (green). We also show the possible detection limit for the proposed Einstein Telescope: {\tt einstein} (orange). For each detector, we show results for three values of the magnification: $\mu=1$ (solid), $\mu=3$ (dashed), and $\mu=10$ (dotted). The cyan dot shows the parameters for GW150914.}
\label{fig:SNRcontour}
\end{figure}

For a given detector $\mathcal{D}$, a GW event magnified by a factor of $\mu$ has a matched-filtering signal-to-noise (SNR) ratio
\ba
\label{eq:SNR-mu}
\mathcal{S}_{\mathcal D} \left( M, z, \mu \right) \equiv \mathcal{S}_{\mathcal D} ( \tilde M, \tilde z ) = \sqrt{\mu}\,\mathcal{S}_{\mathcal D}\left(M,z\right),
\ea
where $\mathcal{S}_{\mathcal D}\left(M,z\right)$ is the SNR without lensing. We assume that events with SNR lower than some threshold value are not detected. Throughout, we use the conventional threshold value for a single detector $S_0=8$ \cite{2015CQGra..32a5014M,2014arXiv1409.0522C}. Hence, another effect of large magnifications is to push undetectable or marginally detectable mergers above the (fixed) detection threshold. Equation \eqref{eq:SNR-mu} implies that lensing magnification extends a given detector's redshift reach. Figure \ref{fig:SNRcontour} demonstrates this effect for several detectors. To simplify our discussion, we only consider equal-mass mergers, and always compute signal-to-noise ratios averaged over all orbital orientations. Note also that if a massive merger occurs at a sufficiently high redshift, the signal tends to redshift out of the detectors' frequency band. This adversely impacts detectability, as illustrated by the turnover of the threshold curves toward high masses.

\section{Magnification probability}
\label{sec:mpdf}

We can model compact binary mergers as point sources to an excellent degree of accuracy. If we assume a uniform distribution of binary mergers on the {\it source plane} at a source redshift $z$, a random merger has a probability
\ba
dP & = & \frac{dP(\mu;z)}{d\ln\mu}\,d\ln\mu.
\ea 
of being magnified by a factor close to $\mu$, where the factor multiplying $d\ln\mu$ on the RHS is a probability density function (PDF). This satisfies the usual normalization condition when integrated over all possible values of $\mu$.  To a very good approximation, the mean magnification is unity, i.e., the total solid angle is conserved~\cite{Weinberg1976lensing,Kaiser:2015iia}. Mathematically,
\ba
\label{eq:muPDF-mean}
\VEV{\mu} \equiv \int^{+\infty}_0\,d\ln\mu\,\frac{dP(\mu;z)}{d\ln\mu}\,\mu & = & 1.
\ea  
The PDF in the weak lensing regime has been studied by Refs. ~\cite{Munshi:1999qw,Wang:2002qc,Das:2005yb,Amendola:2013twa}.Those do not include strong lensing by isolated virialized clumps, which is responsible for significantly biasing the mass and distance estimates. 

Previous studies used ray-tracing through simulated large-scale structure to study the full magnification PDF covering both the weak and strong lensing regimes. The results exhibit an asymptotic power-law tail at large magnifications, i.e., $dP/d\mu \sim \mu^{-3}$. This tail is a generic feature of the point source being located near the fold caustics of a single lens plane \cite{1992grle.book.....S}. Moreover, the tail's amplitude grows rapidly with source redshift~\cite{Hilbert:2007ny,Takahashi:2011qd}. The stellar component located in galactic cores can also greatly enhance the strong magnification probability when compared to the case of dark-matter-only halos~\cite{Hilbert:2007jd}. The resolution of these studies does not permit the inclusion of microlensing by stars. However, we anticipate that microlensing is unimportant for GW magnification since the relevant wavelengths are larger than typical stellar Schwarzschild radii~\cite{Takahashi:2003ix}. In any case, galaxy lenses' contribution to the strong lensing optical depth dominates that of stellar microlenses for sources at cosmological distances. 

\begin{figure}[t]
\centering
\hspace{-0.5cm}
\includegraphics[scale=0.65]{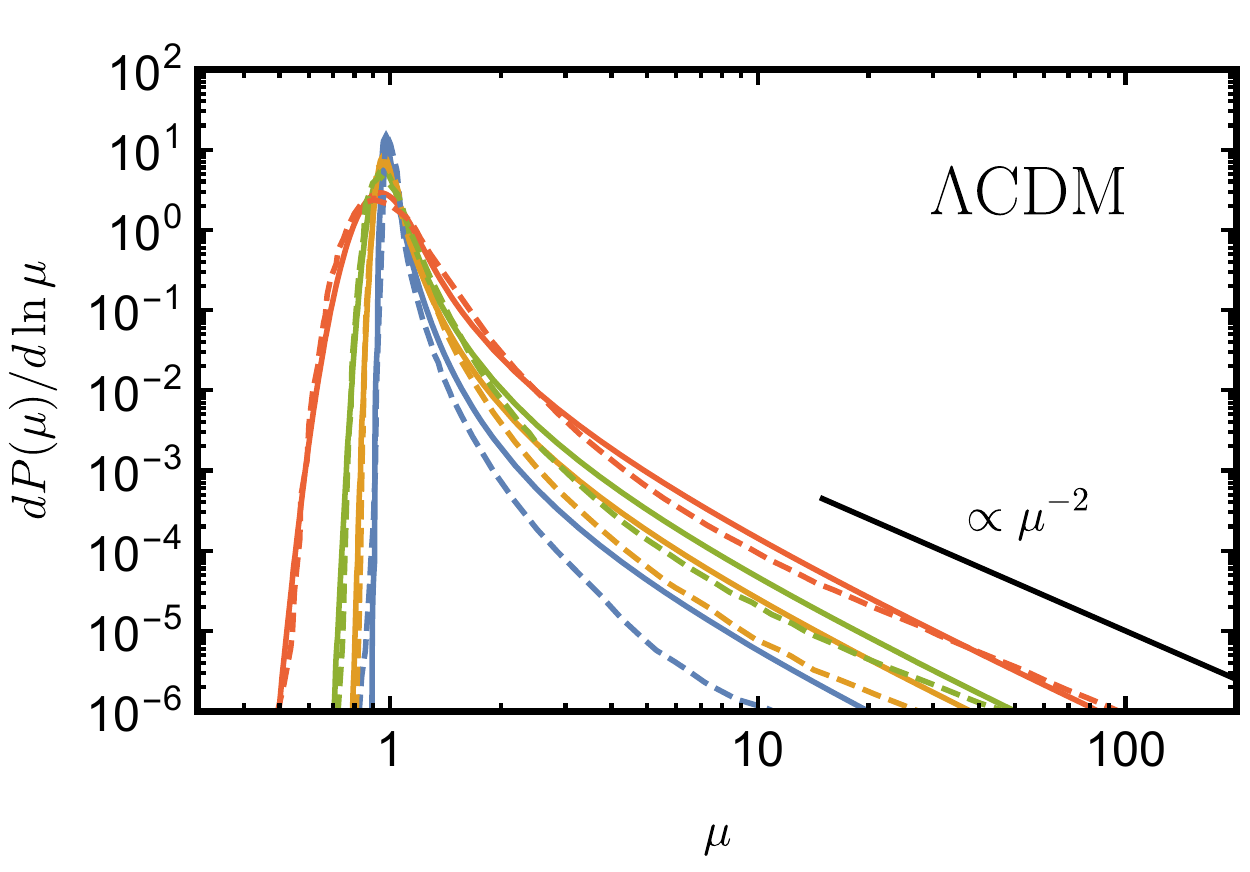}
\caption[]{Magnification probability distribution $dP/d\ln\mu$ produced by parametric fit \refeq{semi-analytical-dai}. Point source is assumed. From bottom to top, respectively, we show a number of representative source redshifts $z=1,\,2,\,3,\,10$. PDFs measured by Ref.~\cite{Takahashi:2011qd} from dark-matter-only simulation are over-plotted in dashed curves, compared to which our fit accounts for enhancement of strong lensing by baryonic components according to Ref.~\cite{Hilbert:2007jd}.}
\label{fig:muPDF}
\end{figure}

Instead of performing ray-tracing simulations, we provide a recipe to fit the source magnification PDF (\reffig{muPDF}), which allows for efficient computation. Appendix \ref{app:lenPDFfitLiang} details our fitting formula, which fits the weak lensing part around $\mu = 1$ to Figure 7 of Ref.~\cite{Takahashi:2011qd} for each source redshift $z$, and matches the high magnification tail onto the optical depth for $\mu>10$ from Figure 2 of Ref.~\cite{Hilbert:2007jd} (the latter includes the contribution of stellar mass.). Our method leads to higher probability for strong lensing $\mu>2$ than Ref.~\cite{Takahashi:2011qd} (a few times increase for $z>2$ and even larger for $z<2$). However, it is unlikely to be a significant overestimation of the reality, as the large $\mu$ tail is calibrated to the baryonic enhancement found by Ref.~\cite{Hilbert:2007jd}.

It should not be taken for granted that our fit gives the true lensing rate with high accuracy for all magnifications. While our fit tracks the power law tail and precisely reproduces the $\mu>10$ optical depth of Ref.~\cite{Hilbert:2007jd}, it differs from the numerical results of Ref.~\cite{Takahashi:2011qd} at the $40\%$ level in the weak lensing regime, i.e., $0.9<\mu<1.1$ for $z>2$, and has even larger uncertainties at lower source redshifts. Besides, our fit can overestimate the probability in the highly demagnified part of the distribution. Nevertheless, since appreciable biases in the mass and redshift estimates arise only due to large magnifications, i.e., $\mu \gtrsim 3$ (see \reffig{ztotildez}), uncertainties in the weak-lensing magnification do not impact our main conclusions. We note that our formula is properly normalized, and self-consistently satisfies \refeq{muPDF-mean}. As a check, we also fit the large-magnification tails to the results of Ref.~\cite{Takahashi:2011qd}, which are systematically lower than those of Ref.~\cite{Hilbert:2007jd} due to their neglect of the stellar contribution, and found that our major results were indistinguishable from those obtained by directly applying the numerical PDFs of Ref.~\cite{Takahashi:2011qd}.

Extremely large magnification factors arise when the source approaches a caustic of the lens mapping, where it produces a close pair of images (the magnification is formally divergent at the caustic). At such locations, the effects of diffraction smear out the lens mapping~\cite{1983ApJ...271..551O} and terminate the power-law tail in the magnification PDF. Appendix \ref{app:waveeffects} shows that this truncation occurs at $\mu \gtrsim 10^3$. Given the minuscule probability for such high values, we can safely ignore this effect in the rest of our calculations. 

\section{Statistical effects of lensing}
\label{sec:statistics}

Astrophysical models of stellar binary evolution predict the differential rate density $d^2n(M,z)/(dM\,dt_s)$, namely the BH-BH merger rate per unit comoving volume and unit proper time $t_{\rm s}$ at redshift $z$ and at an intrinsic mass scale $M$.\footnote{We primarily discuss BH-BH binary mergers within the band of ground-based detectors. For simplicity, we only consider equal-mass mergers with $M$ being the mass of either binary component. Note that extreme values of mass ratio degrade the signal-to-noise ratio.} We briefly outline the procedure to convert this rate density into an observed one, and study the impact of lensing and detector sensitivity.

\begin{figure}[t]
\centering
\hspace{-0.5cm}
\includegraphics[scale=0.65]{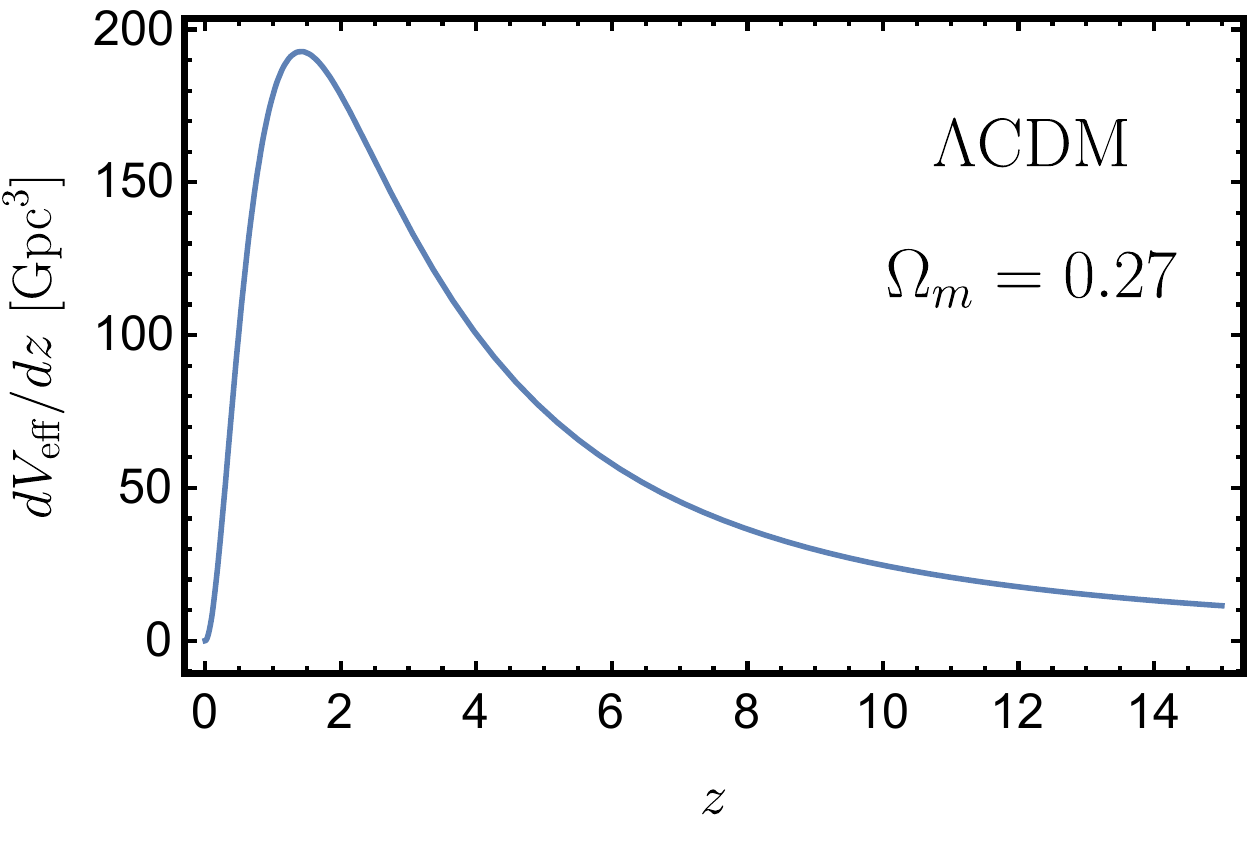}
\caption[]{The redshift distribution of effective comoving volume $dV_{\rm eff}(z)/dz = 4\pi \,c\, \chi^2(z)/((1+z)\,H_0\,E(z))$ for the fiducial $\Lambda$CDM cosmology.}
\label{fig:dVeffdz}
\end{figure}

First, let us ignore the effects of lensing. A GW detector detects mergers from any redshift $z$ along its past light cone, measured with respect to its local proper time $t$. Proper time intervals at the detector and source redshifts are related by $dt = (1+z)\,dt_s$. We use the expression for the comoving volume in a flat $\Lambda$CDM cosmology to obtain an {\em observed} differential rate
\ba
\label{eq:d3NdMdzdt}
\frac{d^3 N(M,z)}{dM\,dz\,dt} & = & \frac{d^2n(M,z)}{dM\,dt_s}\,\frac{4\pi\,c\, \chi^2(z) }{(1+z)\,H_0 \,E(z)},
\ea
where $c$ is the speed of light, $\chi(z)$ is the comoving distance out to redshift $z$, and $E(z)$ is the Hubble expansion rate at redshift $z$ in units of the current value, $H_0$. For a flat $\Lambda$CDM cosmology
\ba
E(z) & = & \sqrt{\Omega_m(1+z)^3+1-\Omega_m}, \ {\rm and} \\
\chi(z) & = & \frac{c}{H_0}\,\int^z_0\,\frac{dz'}{E(z')}.
\ea
In a homogenous cosmology, the comoving distance $\chi(z)$ is related to the luminosity distance $d_L(z)$ by $\chi(z) = d_L(z)/(1+z)$. The factor of $4\pi \,c\,\chi^2(z)/((1+z)\,H_0\,E(z))$ in \refeq{d3NdMdzdt} is an effective comoving volume per unit redshift interval, which connects the intrinsic and observed rate densities.\footnote{This is related to the comoving colume element of Ref.~\cite{1999astro.ph..5116H}, with an extra factor accounting for the change in proper time.} With increasing source redshift $z$, a larger comoving distance permits access to a greater volume; on the other hand, redshifting of the source-frame rate $\propto (1+z)^{-1}$ decreases the effective volume per unit proper time at the observer. As \reffig{dVeffdz} shows, the net result is that a major fraction of the observed events within the horizon would occur around $z\sim 1-2$ in the absence of any intrinsic redshift evolution, with a total available effective comoving volume of a few hundred cubic giga-parsec.

Let us now add a finite probability for large lensing magnification. As we discussed in Sec.~\ref{sec:degeneracy}, we cannot access the mergers' intrinsic parameters, and therefore the direct observable is the differential rate with respect to the inferred quantities $\tilde M$ and $\tilde z$,
\ba
\label{eq:obs-rate-lensing}
\frac{d^3 N (\tilde M,\tilde z)}{d\tilde M\,d\tilde z\,dt} &=&  \int\, d\ln\mu\,\frac{dP(\mu;z)}{d\ln\mu} \en 
&& \times \frac{d^3 N (M,z)}{dM\,dz\,dt}\,\left|\frac{\partial(M,z)}{\partial(\tilde M,\tilde z)} \right|_\mu.
\ea
The conversion between the intrinsic and extrinsic parameters leads to a factor of the Jacobian of the map defined by Eqs.~\eqref{eq:Mzdegen} and \eqref{eq:mudLdegen}, which evaluates to
\ba
\left|\frac{\partial(M,z)}{\partial(\tilde M,\tilde z)} \right|_\mu & = & \frac{(1+\tilde z)\,d'_L(\tilde z)}{(1+z)\,d'_L(z)} \frac{d_L(z)}{d_L(\tilde z)},
\ea
where the primes indicate differentiation with respect to the arguments.  

We can compute a cumulative event rate by integrating \refeq{obs-rate-lensing} further over redshift $\tilde z$ and mass $\tilde M$. For a detector/detector-network $\mathcal{D}$, the observed cumulative rate for BHs that appear heavier than $M_{\rm min}$ is
\ba
\label{eq:obs-rate-lensing-cumu}
&& \frac{d N_{\cal{D}}}{dt}\left(\tilde M > M_{\rm min}\right) = \int^{+\infty}_{M_{\rm min}} d\tilde M \int d\tilde z\, \Theta\left(\mathcal{S}_{\cal{D}}(\tilde M,\tilde z)-\mathcal{S}_0\right) \en 
&& \times \int\, d\ln\mu\,\frac{dP(\mu;z)}{d\ln\mu} \frac{d^3 N (M,z)}{dM\,dz\,dt}\,\left|\frac{\partial(M,z)}{\partial(\tilde M,\tilde z)} \right|_\mu.
\ea
Note that the above equation includes a detectability cut through its requirement that the signal-to-noise ratio $\mathcal{S}_{\cal{D}}(\tilde M, \tilde z)$ exceeds a threshold $\mathcal{S}_0$. This form has the advantage that the detector-specific SNR $\mathcal{S}_{\cal{D}}(\tilde M,\tilde z)$ is disentangled from the intrinsic merger rate and the magnification PDF, since it only depends on the inferred parameters $\tilde M$ and $\tilde z$. A practical consequence is that while different detectors probe different areas in the $(\tilde M, \tilde z)$-plane, the lensed fraction of any given $(\tilde M, \tilde z)$ bin is independent of the detector.

The above formalism is applicable to any general differential rate density of BH-BH mergers, $d^2n(M,z)/(dM\,dt_s)$. Given our sparse knowledge of this quantity,  we do not attempt to suggest a precise rate that corresponds to reality. Rather, our strategy is to survey a variety of possibilities with qualitatively different behaviors, and to identify in each case the possible effects of lensing magnification on the observed differential rate.

Before we consider specific models of the rate density in the literature, we will demonstrate some features of the cumulative rate in \refeq{obs-rate-lensing-cumu}, both generally and within a toy model for the evolution of the merger rate that we tune to emphasize the effects of lensing.

\begin{figure}[t]
\centering
\hspace{-0.5cm}
\includegraphics[scale=0.5]{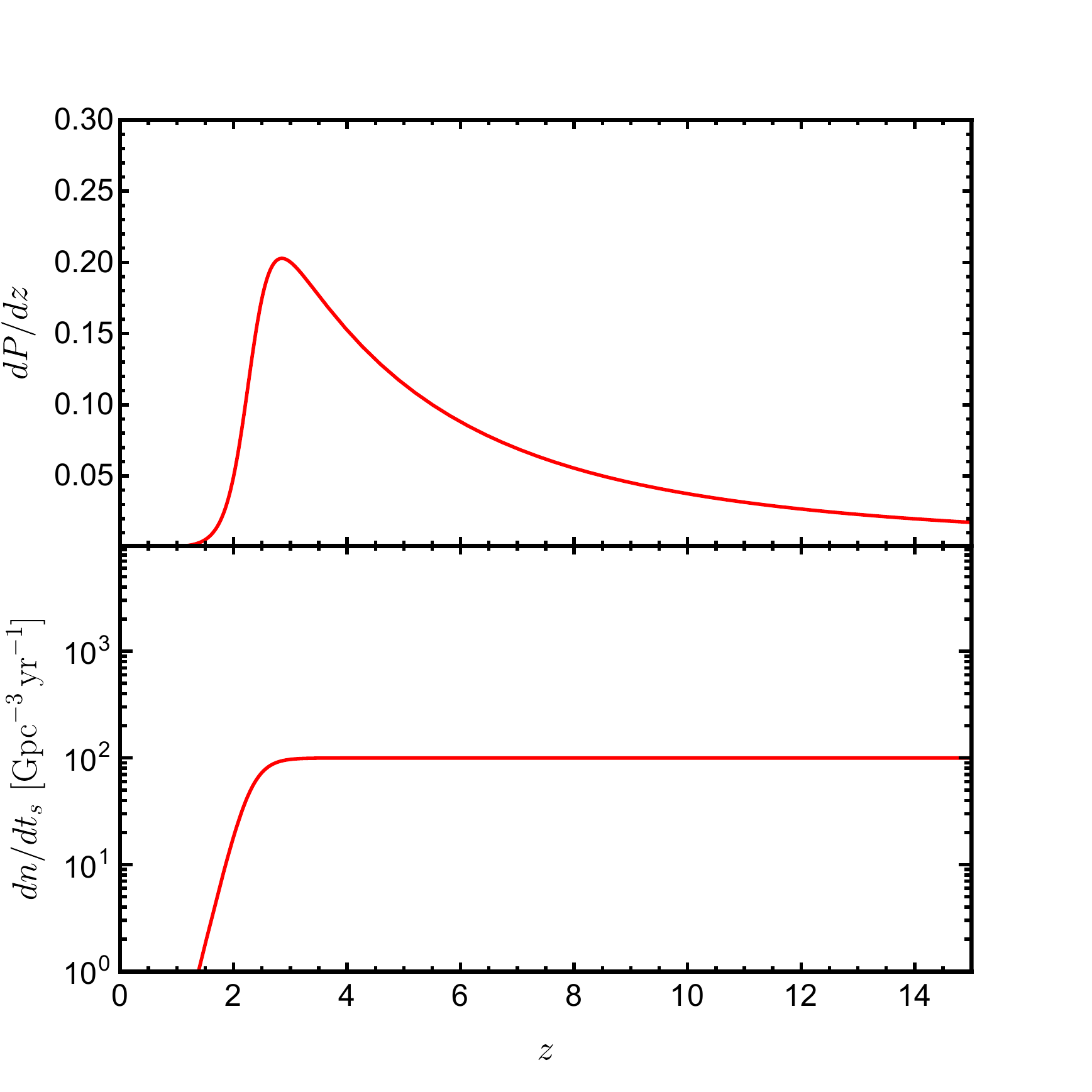} \\
\caption[]{(Physical) redshift distribution (normalized) of all BH-BH mergers on the observer's past light cone ({\it upper panel}), and the redshift evolution of the source-frame merger rate density ({\it lower panel}), for the illustrative model of high-$z$ mergers we consider in \refsec{statistics}.}
\label{fig:pop3new_zdist}
\end{figure}

Consider a case where the intrinsic rate $d^2n(M,z)/dM\,dt_s$ varies mildly with redshift, but is sharply cut off beyond a certain mass $M_{\rm max}$. Mergers with higher masses, $\tilde M > M_{\rm max}$, can still show up due to strong lensing. Suppose we can approximate the luminosity distance by a power law $d_L(z) \propto (1+z)^\gamma$ (valid at cosmological redshifts). The observed differential rate for heavy mergers, i.e., those with $\tilde M > M_{\rm max}$, is roughly
\ba
\hspace{-0.5cm}
\frac{d^2N}{d\tilde M\,dt_s} & \sim  & \int\,d\ln\mu\,\frac{dP(\mu;z)}{d\ln\mu}\,\frac{d^3N}{dM\,dz\,dt}\,\left|\frac{\partial(M,z)}{\partial(\tilde M,\tilde z)} \right|_\mu.
\ea

\begin{figure*}[t]
\centering

\hspace{-0.5cm}
\includegraphics[scale=0.37]{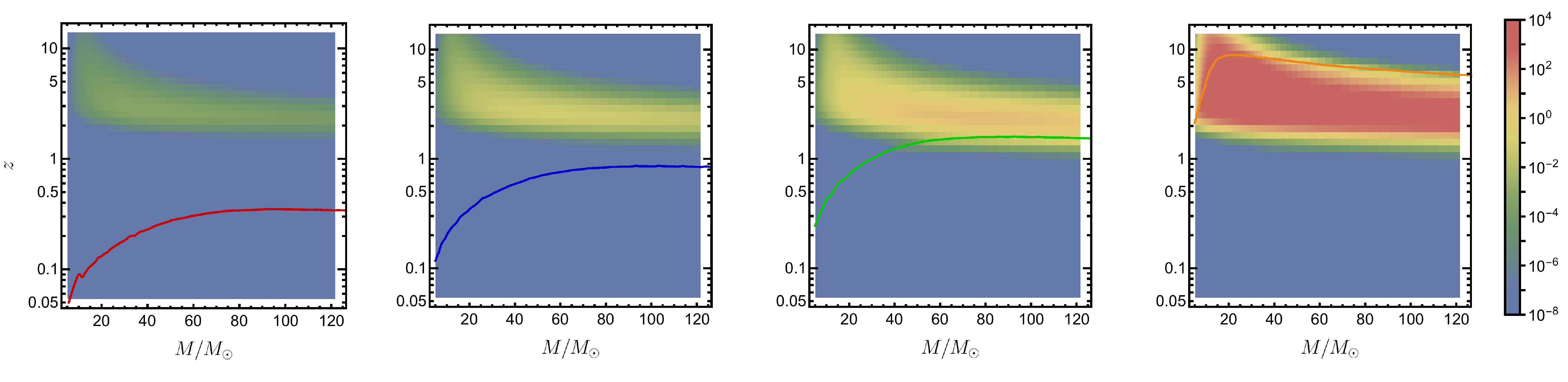}\\

\caption[]{Distribution of true mass $M$ and true redshift $z$ for detectable BH-BH mergers $d^3N/(dt\,d\ln M\,d\ln z)\, [{\rm yr}^{-1}]$ for the illustrative model of high-$z$ mergers discussed in \refsec{statistics}. From left to right, unlensed detection thresholds for four different sensitivities are over-plotted with the same color coding of \reffig{SNRcontour}.}
\label{fig:detect_rate_pop3new}
\end{figure*}

\begin{figure*}[t]
\centering

\vspace{0.7cm}
\hspace{-0.2cm}
\begin{minipage}[]{0.48\linewidth}
\includegraphics[scale=0.43]{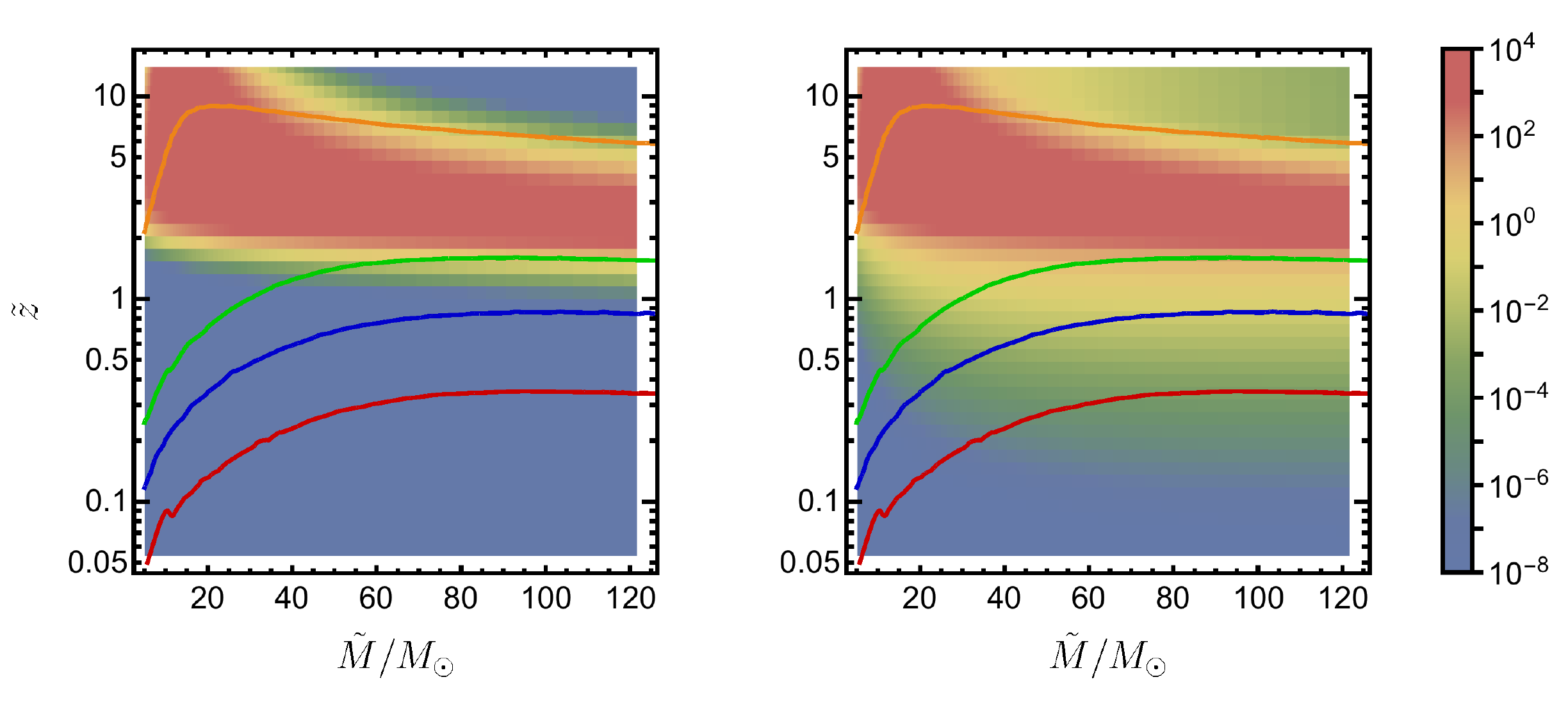}
\end{minipage}
\vspace{-1cm}
\hspace{0.6cm}
\begin{minipage}[]{0.48\linewidth}
\includegraphics[scale=0.43]{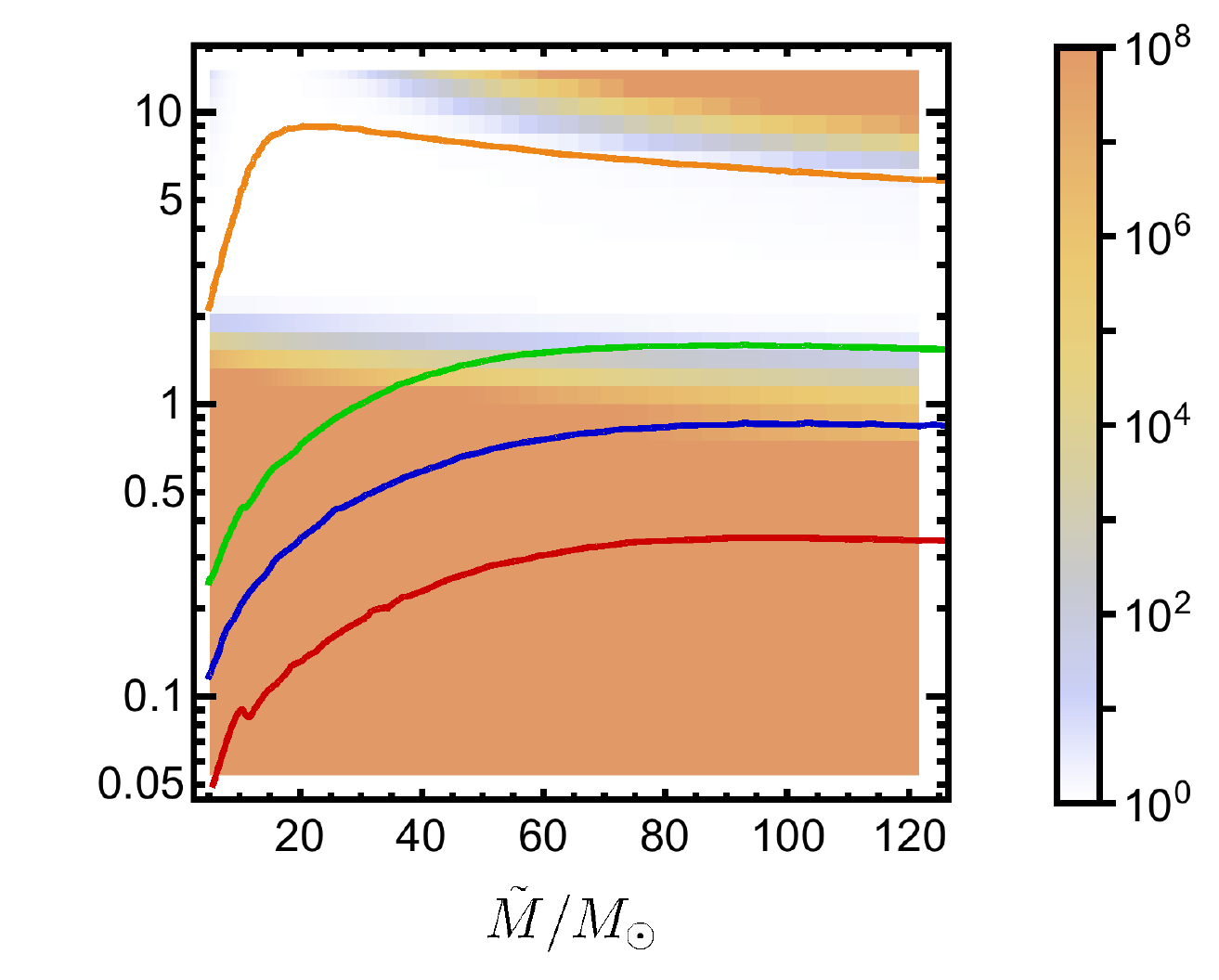}
\end{minipage}\\

\vspace{1cm}
\caption[]{Distribution of inferred mass $\tilde M$ and inferred redshift $\tilde z$ for BH-BH mergers $d^3N/(dt\,d\ln\tilde M\,d\ln\tilde z)\,[{\rm yr}^{-1}]$ for the illustrative model of high-$z$ mergers we consider in \refsec{statistics}. The case with lensing ({\it Middle}) is compared to the case without lensing ({\it Left}). The rate enhancement by lensing is also shown ({\it Right}). Each of the four detectors considered in \reffig{SNRcontour} cuts off the region according to the threshold curves over-plotted, with the same color coding adopted as before.}
\label{fig:obs_rate_pop3new}
\end{figure*} 

Suppose further that the relevant magnifications are large enough that $dP/d\ln\mu \propto \mu^{-2}$. The Jacobian factor may be simply approximated as $M/\tilde M$ at fixed $\mu$,
\ba
\left( M/\tilde M \right)_\mu \sim \frac{1+\tilde z}{1+z} \sim \left( \frac{d_L(\tilde z)}{d_L(z)} \right)^{1/\gamma} \sim \mu^{-1/(2\gamma)}.
\ea
Therefore, the observed differential rate scales as
\ba
\frac{d^2N}{d\tilde M\,dt_s} \propto \mu^{-2 - 1/(2\gamma)} \propto \left( M/\tilde M \right)^{4\gamma+1}.
\ea 
In the above estimate, we substitute for $M$ the maximum mass where the intrinsic rate cuts off. This crude estimate suggests a power-law tail of apparently massive events with an index $4\gamma+1$. For redshifts in the range $3<z<8$, the power law exponent of the luminosity distance is $\gamma \approx 1.5$, and hence the cumulative count $dN(>\tilde M)/dt$ decays as $\tilde M^{-6}$. In any practical scenario, the intrinsic merger rate is redshift dependent and the detectors have a threshold sensitivity. Therefore the observed differential rate departs from the simple power-law that we derived above.

We now consider a toy model within which we relax the above simplifying assumptions. In this model, a copious number of stellar-mass BH binaries merge efficiently enough at high redshifts so that the majority of their mass is in single BHs by $z \simeq 2$. At lower redshifts, the merger rates are low enough that current ground-based detectors do not see a significant number of unlensed events.

We express the intrinsic differential merger rate density as
\ba
\label{eq:d2ndMdts-hier}
\frac{d^2n(M,z)}{dM\,dt_s} & = & \frac{dn(z)}{dt_s}\,\frac{dP(M;z)}{dM},
\ea
where $dn(z)/dt_s$ is the normalization, and $dP(M;z)/dM$ is the probability distribution for the merger mass $M$ at redshift $z$. We assume that the normalization starts out at a large value at $z=20$, and remains constant up to $z \approx z_{\rm end}=2.3$, when merger activity dies off gradually over a half-width of $\Delta z \sim 0.4$. We use the form
\ba
\label{eq:dndt}
\frac{dn(z)}{dt_s} & = & \left(\frac{dn}{dt_s} \right)_0\,\frac12 \left[ 1 + \tanh\left(\frac{z-z_{\rm end}}{\Delta z}\right) \right].
\ea
We choose an initial normalization $(dn/dt_s)_0 = 100\,{\rm Gpc}^{-3}\,{\rm yr}^{-1}$. 

The probability distribution function $dP(M;z)/dM$ is more complicated. We initialize it to a log-normal distribution with a peak at $M=12\,M_\odot$, and a width $\sim 6\,M_\odot$ at $z=20$ (this translates into a small mass fraction $\Omega_{BH}\,\simeq 3 \times 10^{-8}$ residing in initial BHs). We use a merger tree to model the distribution's evolution through a series of hierarchical mergers, i.e., we randomly and repeatedly replace two existing BHs with one new BH (with the appropriate mass).\footnote{Note that late-time mergers involve very massive BHs, and hence are out of the frequency band of ground-based observatories.} In general, the merging BHs have unequal masses, but the mass-ratio distribution does not have significant weight at extremely small values (say, $M_</M_> \lesssim 0.1$). We simplify our analysis by assuming equal-mass mergers with either component containing half the total mass. We do not model the inspiral timescale per merger, but rather assume that the overall merger rate is statistically given by \refeq{dndt}. Note that such a hierarchical merger history is typical of scenarios of Pop III remnants~\cite{Madau:2001sc,Pelupessy:2007mt}, where BHs concentrate in the cores of galaxies due to assembly or dynamical friction. 

We find that the resulting PDF for merger masses can be well fit by a log-normal distribution that shifts toward larger masses with decreasing redshift, i.e.,
\ba
\hspace{-0.7cm} \frac{dP(M;z)}{dM} & = & \frac{1}{\sqrt{2\pi}\,\sigma(z)\,M}\,\exp\left[ - \frac{\left[\ln(M/\bar{M}(z)) \right]^2}{2\sigma^2(z)} \right],
\ea
with $\sigma(z)$ and $\bar{M}(z)$ measured from the simulated hierarchical merger tree. The upper and lower panels in \reffig{pop3new_zdist} show the redshift distribution of all BH mergers on the observer's past light cone (i.e. all mergers detectable by a perfectly noiseless detector) and the normalization of the rate density, respectively.

We next incorporate a detector sensitivity and the lensing magnification of Sec.~\ref{sec:mpdf}. As we discussed after \refeq{obs-rate-lensing-cumu}, these inputs pick out the range of observable intrinsic merger parameters. Figure \ref{fig:detect_rate_pop3new} shows the results for the detectors that we considered in Sec.~\ref{sec:degeneracy}. The results show that when the detector has a poor redshift reach (such as {\tt current} LIGO), only highly magnified early mergers are detectable. The number of observed events rises quickly with any improvement to the detector sensitivity (\reffig{detect_rate_pop3new}). As a consequence of the parameter degeneracy that we highlighted in Sec.~\ref{sec:degeneracy}, these mergers would be misinterpreted as heavy systems from the recent Universe (\reffig{obs_rate_pop3new}). Even though this model is purely illustrative, it suggests that if LIGO or its upgrades observe a profusion of high-mass mergers ($M \gtrsim 40\,M_\odot$) and a dearth of low-mass ones ($M \lesssim 20\,M_\odot$), it could be explained by hierarchical merger activity starting from low-mass seeds ($M \simeq 10\,M_\odot$) at very high redshift and ending around $z\sim 1-2$.

\section{Astrophysical models of the black hole binary population}
\label{sec:len-highm-tail}

\begin{figure}[t]
\centering
\hspace{-0.5cm}
\includegraphics[scale=0.5]{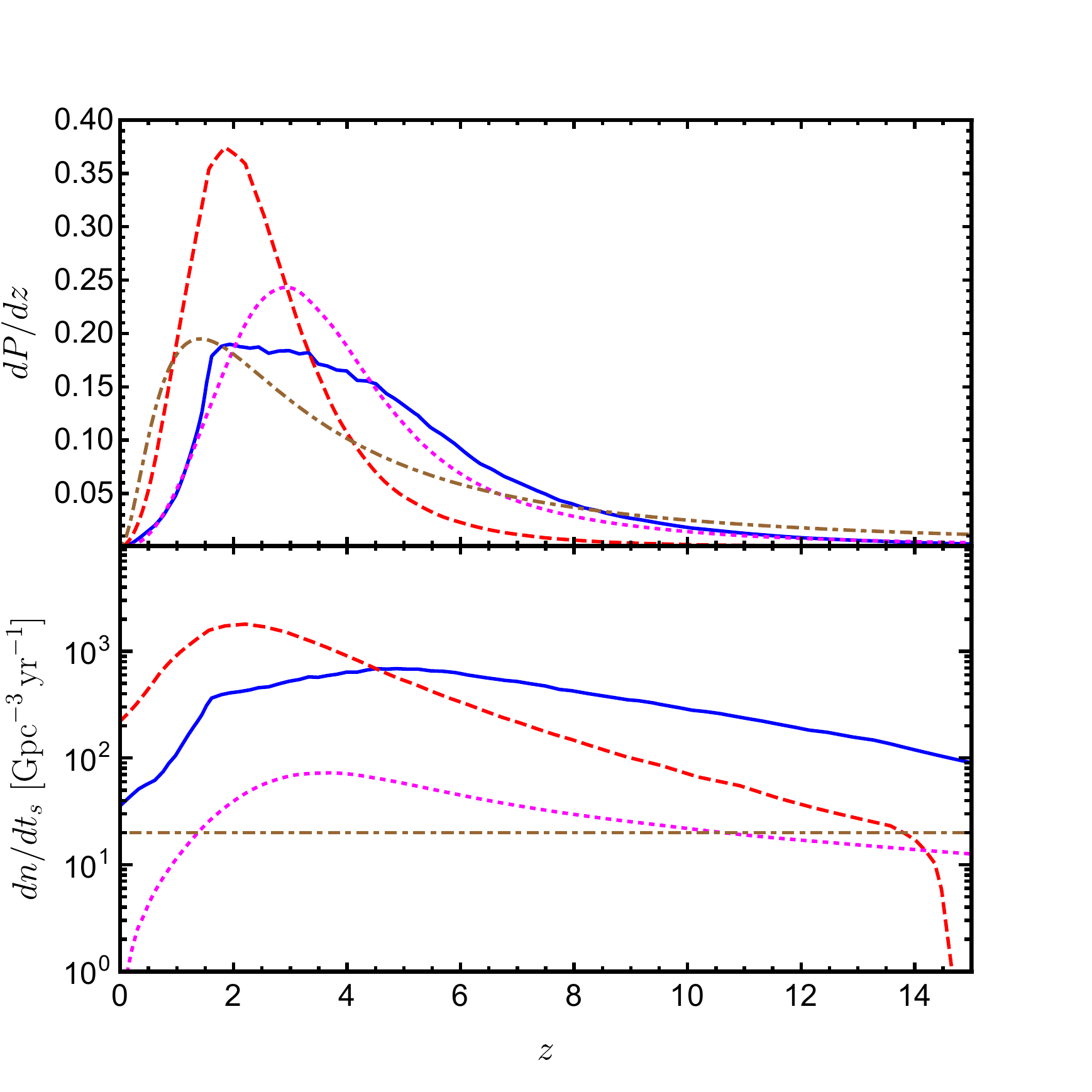} \\
\captionof{figure}[]{(Physical) redshift distribution (normalized) of all BH-BH mergers on the observer's past light cone ({\it upper panel}), and the redshift evolution of the source-frame merger rate density ({\it lower panel}), for different progenitor scenarios we discuss in \refsec{len-highm-tail}: (1) Pop I \& II population synthesis from Ref.~\cite{Dominik:2013tma} (blue solid), and (2) population synthesis from Ref.~\cite{Belczynski:2016obo,Belczynski:2016jno} (red dashed); (3) Hierarchical coalescence of Pop III remnants (magenta dotted); (4) binary mergers from a primordial BH population (brown dash-dotted).}
\label{fig:zdist}
\end{figure}

In this section, we use the methods of Sec.~\ref{sec:statistics} to study the population of lensed BH mergers in several channels of binary BH production that have been proposed in the literature. The exact form of the populations in various channels differ, but when compared to our toy model, all the channels tend to produce a substantial number of mergers in LIGO's band without the help of lensing. Hence, our expectation is that strongly lensed events will contribute to (and in some cases, even dominate) the tail of the observed merger mass distribution, rather than comprise its entirety as in our toy model.  This heavy-mass tail would be analogous to the observed excess of very luminous quasars compared to theoretical expectations~\cite{Narayan:1993gt}.

We consider models with binary black holes that originate from, respectively, the standard populations of Population I and II stars (metal-rich and metal-poor stars, henceforth Pop I and II stars), hypothesized Population III stars (primordial stars that are extremely metal-deficient, henceforth Pop III stars), and a relic primordial population.

\subsection{Mergers from Pop I and II binaries}
\label{sec:schechter}

\begin{figure*}[htbp]
\centering

\hspace{-0.5cm}
\includegraphics[scale=0.37]{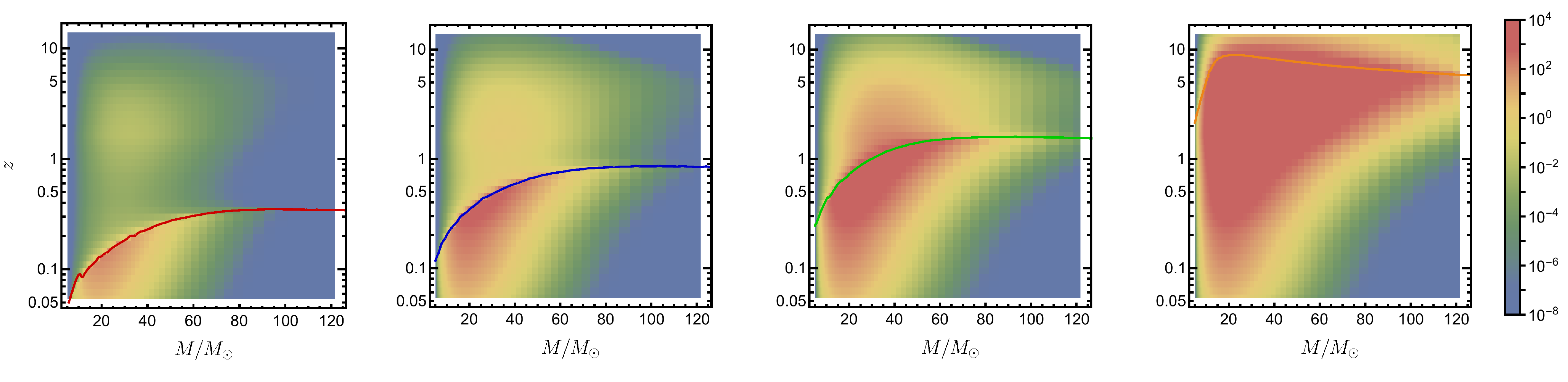}\\

\vspace{-0.3cm}
\hspace{-0.5cm}
\includegraphics[scale=0.37]{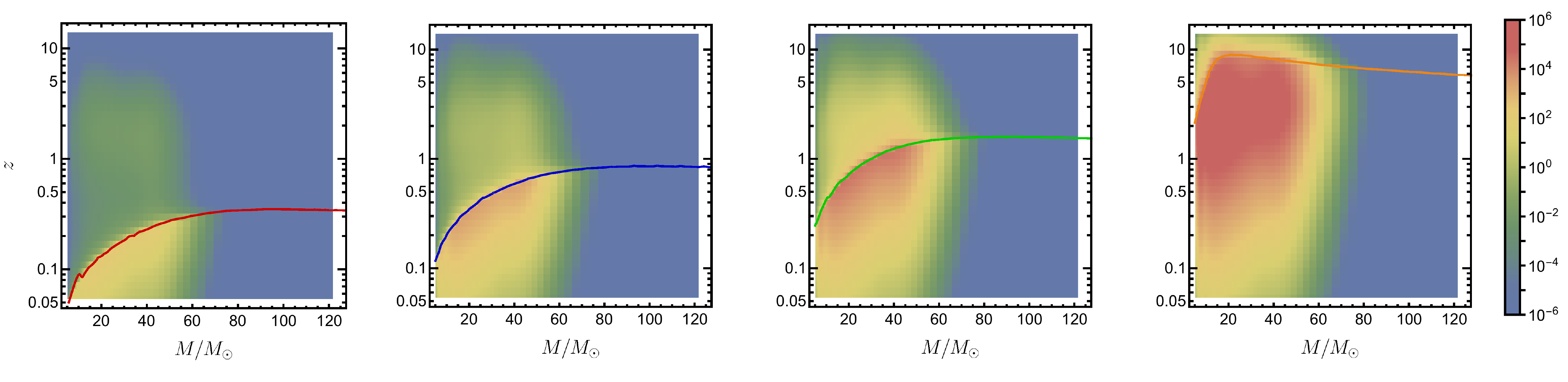}\\

\vspace{-0.3cm}
\hspace{-0.5cm}
\includegraphics[scale=0.37]{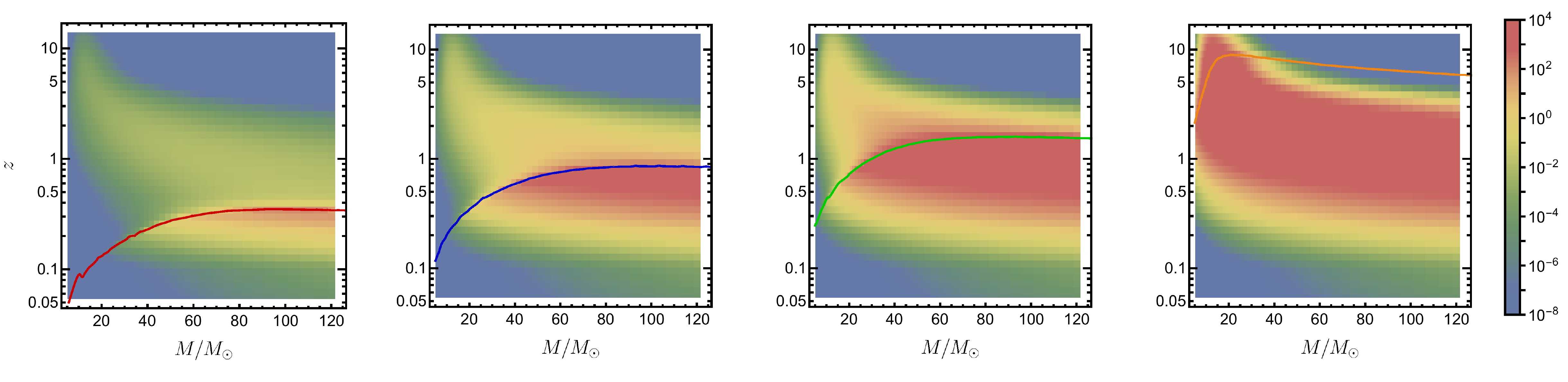}\\

\vspace{-0.3cm}
\hspace{-0.5cm}
\includegraphics[scale=0.37]{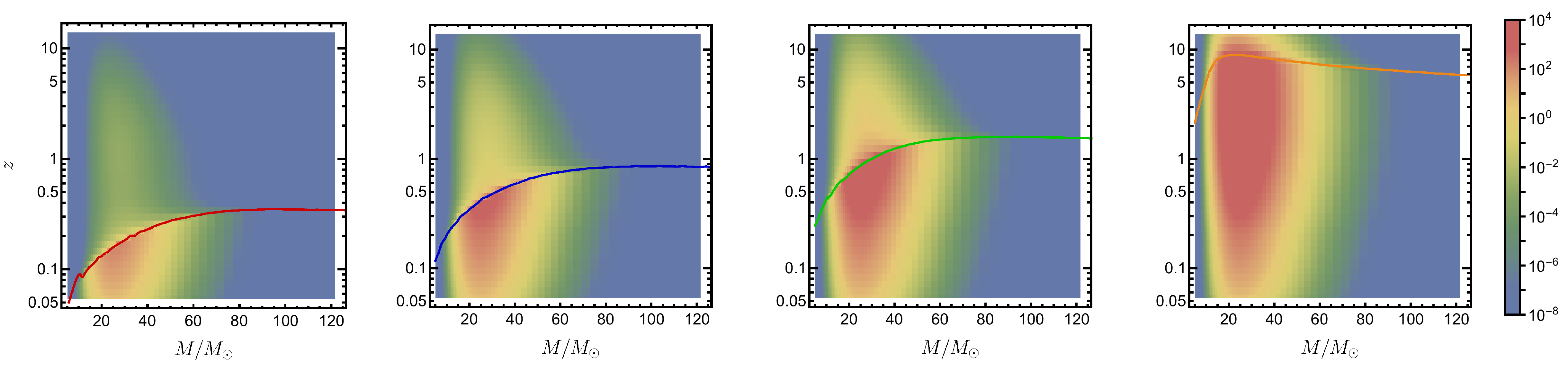}
\caption[]{Distribution of true mass $M$ and true redshift $z$ for detectable BH-BH mergers $d^3N/(dt\,d\ln M\,d\ln z)\, [{\rm yr}^{-1}]$. From top to bottom we plot results for models (1)-(4) in the order they are introduced in \reffig{zdist}. From left to right, unlensed detection thresholds for four different sensitivities are over-plotted with the same color coding of \reffig{SNRcontour}.}
\label{fig:detect_rate}
\end{figure*}

Binaries of Pop I and II massive stars evolving in isolation in low stellar-density environments have been proposed as progenitors for the majority of stellar BH-BH mergers in general, and GW150914 in particular (see e.g. Refs.~\cite{Dominik:2013tma,Belczynski:2016obo}; note however that no BH heavier than $20\,M_\odot$ has been seen in X-ray binaries with reliable mass measurements~\cite{2010ApJ...725.1918O,Farr:2010tu}). In such scenarios, the binary BH merger rate at $z=2-5$ is larger than the local value due to a) the higher star-formation rate, and b) the increased abundance of massive stars due to the lower metallicity in star-forming environments. 

Typical models predict a rapid decay in the merger mass function at the high mass end~\cite{Dominik:2013tma,Belczynski:2016obo}. Below we compare two different models. One is parametrized by intrinsic differential merger rate densities in the form of a Schechter distribution
\ba
\label{eq:schechter_mass_law}
\frac{d^2 n(M,z)}{dM\,dt_s} & = & \frac{dn(z)}{dt_s}\,\frac{\Theta(M-M_{\rm cut})}{M_\star(z)\, \Gamma(1+\gamma(z))}\,\left( \frac{M-M_{\rm cut}}{M_\star(z)} \right)^{\gamma(z)} \en
&& \times \exp\left[ - \frac{M-M_{\rm cut}}{M_\star(z)} \right],
\ea
where $M_{\rm cut}$ and $M_\star(z)$ are cutoff and characteristic masses, respectively. We let the characteristic mass scale $M_\star(z)$ evolve with redshift as
\ba
M_\star(z) & = & 3\,M_\odot\, \left((1+z)/(1+1.5)\right)^{0.5},
\ea 
which gives $M_\star/M_\odot = 1.9, 2.7$, and $3.6$ at $z=0,1$, and $3$ respectively. We apply a low-mass cutoff at $M_{\rm cut}=5\,M_\odot$, and set the power index to $\gamma(z)=6$. With these values, the distribution peaks at $M/M_\odot = 18, 24$, and $32$, with RMS values $\Delta M/M_\odot = 5, 7$, and $10$, at $z=0,1$, and $3$ respectively. These parameters are reasonably compatible with the numerical results of Ref.~\cite{Dominik:2013tma}, and are consistent with our intuition that more massive BHs form from massive stars at high redshifts due to the lower environmental metallicities. 

At each redshift, we normalize the rate density $dn/dt_s(z)$ to match an earlier study Ref.~\cite{Dominik:2013tma}, or the dashed curve of Fig. S5 of Ref.~\cite{Belczynski:2016obo}, which predicts a local merger rate density of $\sim 36\,{\rm Gpc}^{-3}\,{\rm yr}^{-1}$ (compare to the current LIGO-Virgo constraint $2 - 53 \,{\rm Gpc}^{-3}\,{\rm yr}^{-1}$~\cite{Abbott:2016nhf}) and a peak of merger activity at $z \sim 4-6$.

For a second model, we directly use the numerical output for mass distribution and redshift evolution from the latest population synthesis simulation~\cite{Belczynski:2016obo,Belczynski:2016jno} performed with the {\tt StarTrack} code~\cite{2002ApJ...572..407B,2008ApJS..174..223B}.\footnote{For simplicity, we compute the signal-to-noise ratio assuming equal-mass mergers and taking the component mass to be half of the binary mass.} The new simulation has improved upon earlier ones by calibrating to recent observational constraints on the physics of massive binary evolution (see Ref.~\cite{Belczynski:2016jno} for details). In particular, the new simulation assumes that the masses of remnant BHs are  limited to $\sim 50\,M_\odot$ according to improved modeling of the physics involved in pair-instability pulsation supernovae (PPSN)~\cite{2002ApJ...567..532H,2007Natur.450..390W} and pair-instability supernovae (PSN)~\cite{1984ApJ...280..825B,2001ApJ...550..372F,2012ApJ...748...42C}, and that these effects are important in low-metallicity environments $Z<10\%\,Z_\odot$.

In reality, an abrupt cutoff in BH mass in such a complex stochastic process may seem contrived. Scatter in the remnant BH mass is expected when a progenitor star of given mass goes supernova. To mimic such scatter we thus introduce to the simulation data a log-normal random fluctuation with variance $0.04\, {\rm dex}$ to all binary masses, which smoothes the sharp PPSN/PSN cutoff to a plausible width of $\sim 10\,M_\odot$~\cite{BelczynskiPrivateComm}. By manually mitigating the hard cutoff we are conservative about the relative importance of the magnification tail at the high-mass end close to the cutoff threshold. In practice we find this smoothing does not significantly alter the absolute rate or shape of the high-mass tail induced by lensing.

The lower panel of \reffig{zdist} compares the normalization $dn(z)/dt_s$ as a function of redshift between the earlier model of Refs.~\cite{Dominik:2013tma} and the latest results of Ref.~\cite{Belczynski:2016obo,Belczynski:2016jno}. For the latter, a higher local merger rate density $\sim 200 \,{\rm Gpc}^{-3}\,{\rm yr}^{-1}$ is predicted and the peak of merger activities shifts to $z \simeq 2$.

\begin{figure*}[htbp]
\centering

\hspace{-0.2cm}
\begin{minipage}[]{0.48\linewidth}
\includegraphics[scale=0.43]{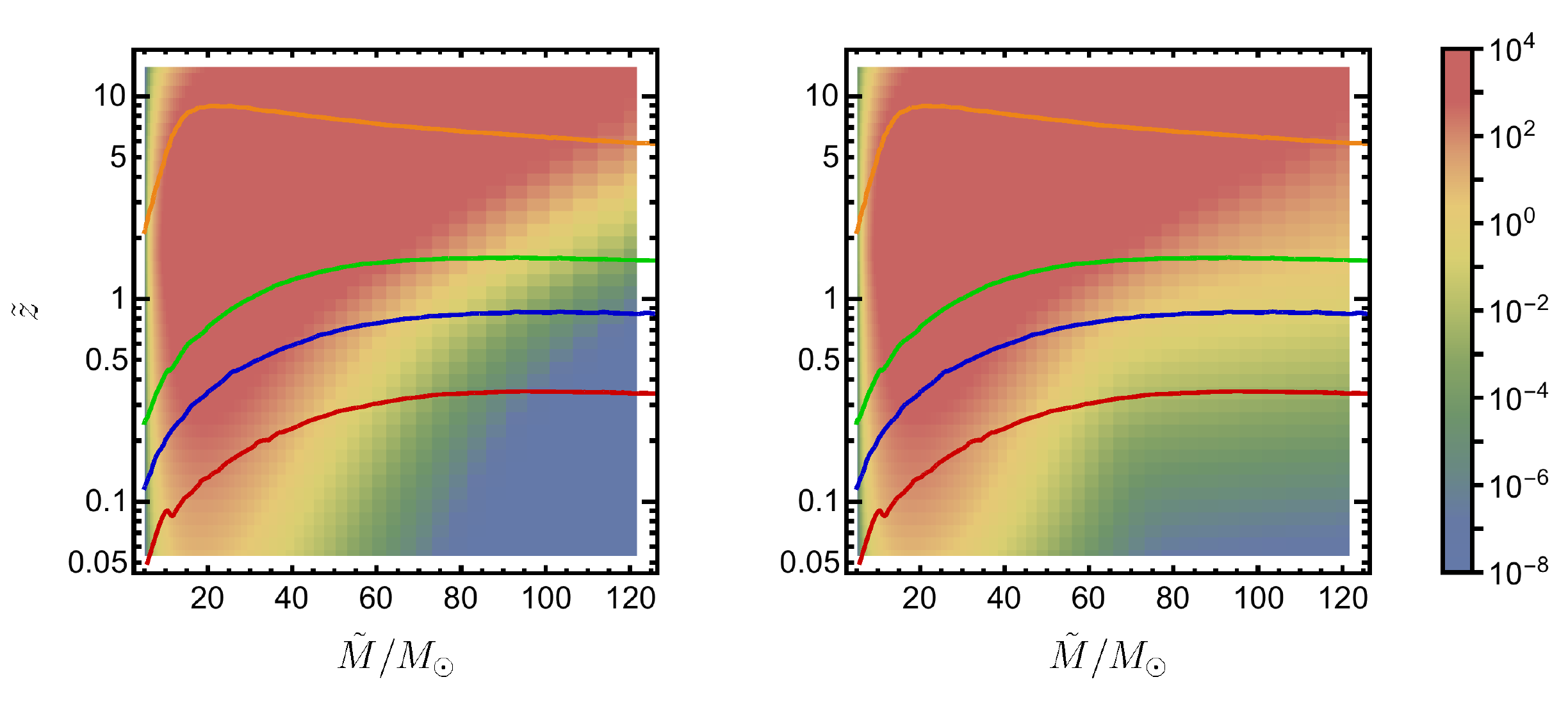}
\end{minipage}
\vspace{-1cm}
\hspace{0.6cm}
\begin{minipage}[]{0.48\linewidth}
\includegraphics[scale=0.43]{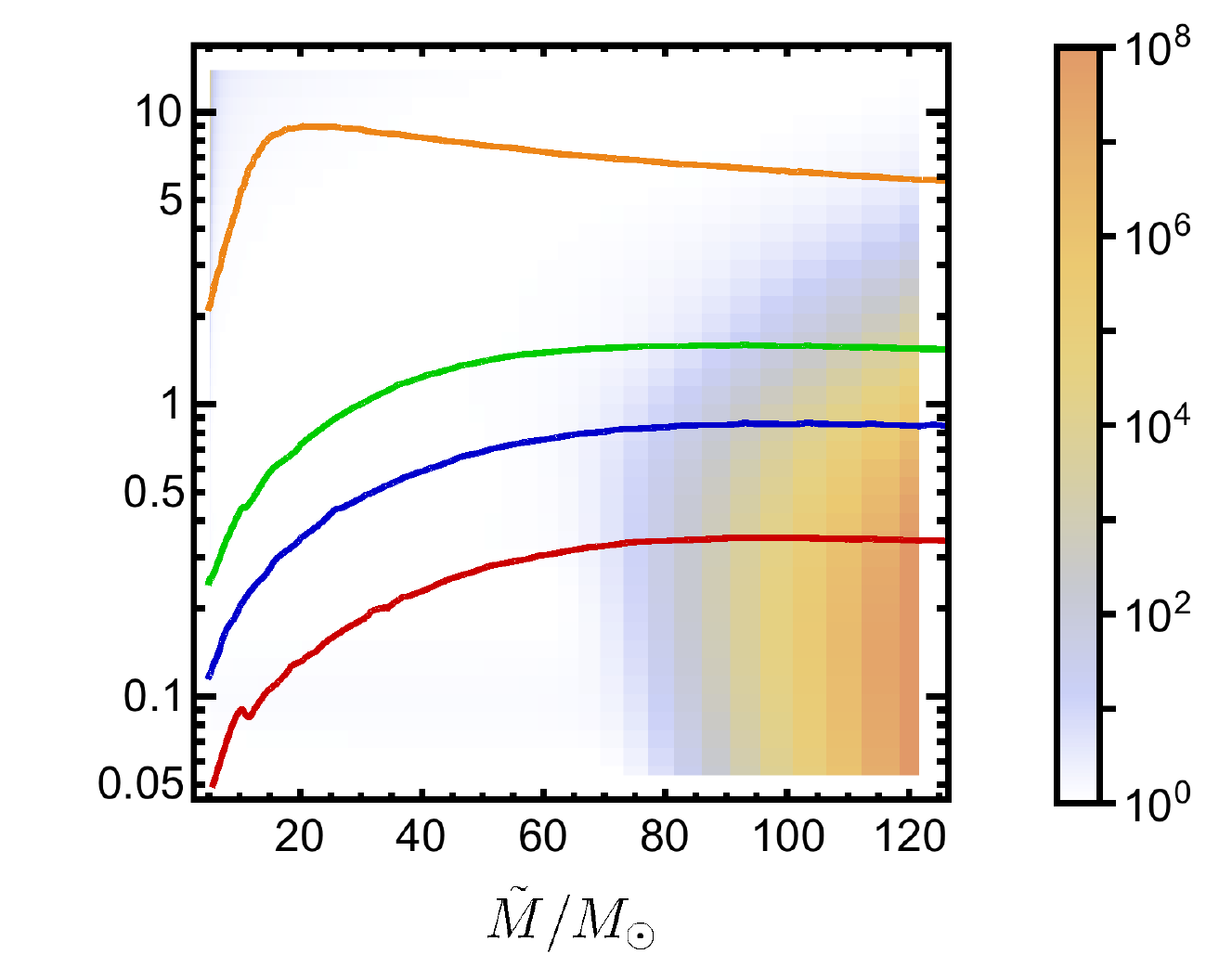}
\end{minipage}\\

\vspace{0.7cm}
\hspace{-0.2cm}
\begin{minipage}[]{0.48\linewidth}
\includegraphics[scale=0.43]{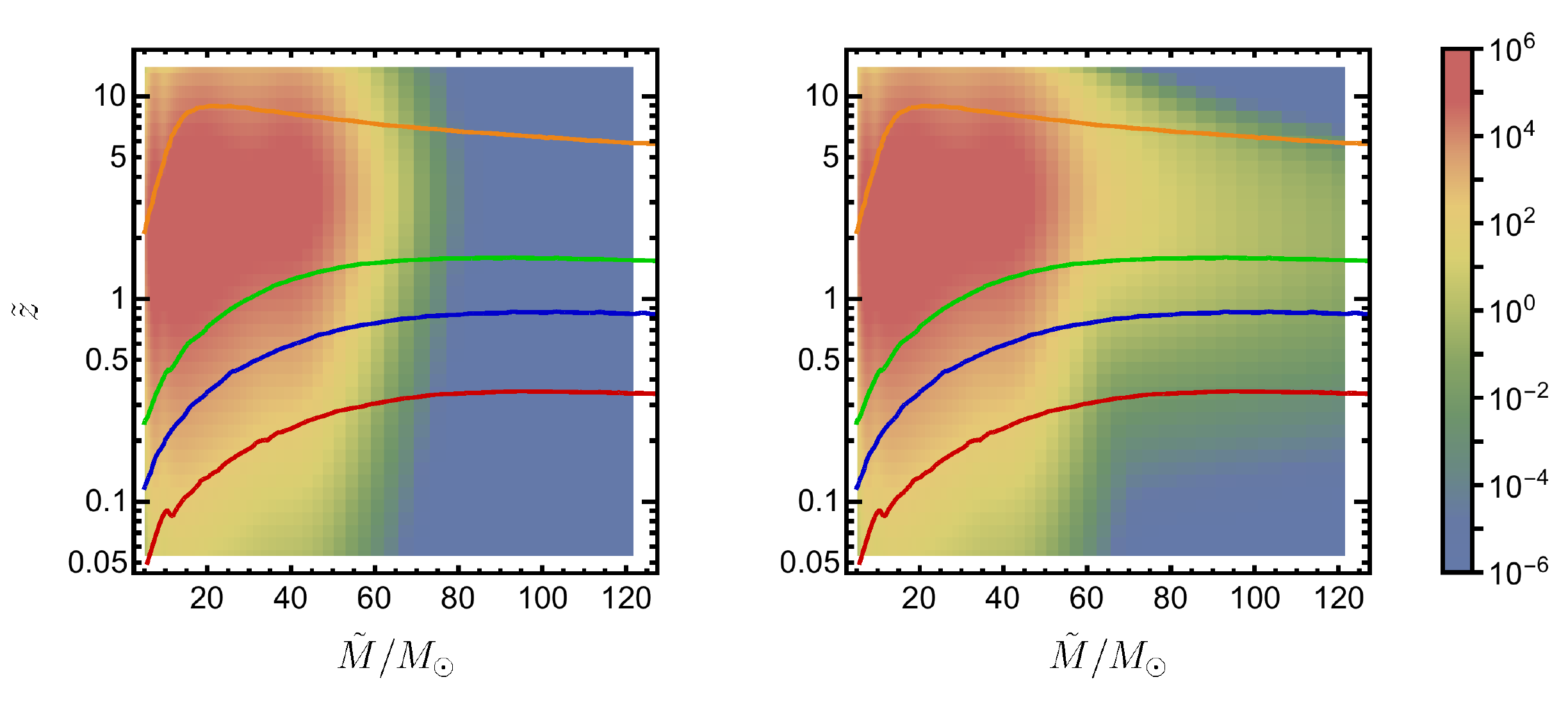}
\end{minipage}
\vspace{-1cm}
\hspace{0.6cm}
\begin{minipage}[]{0.48\linewidth}
\includegraphics[scale=0.43]{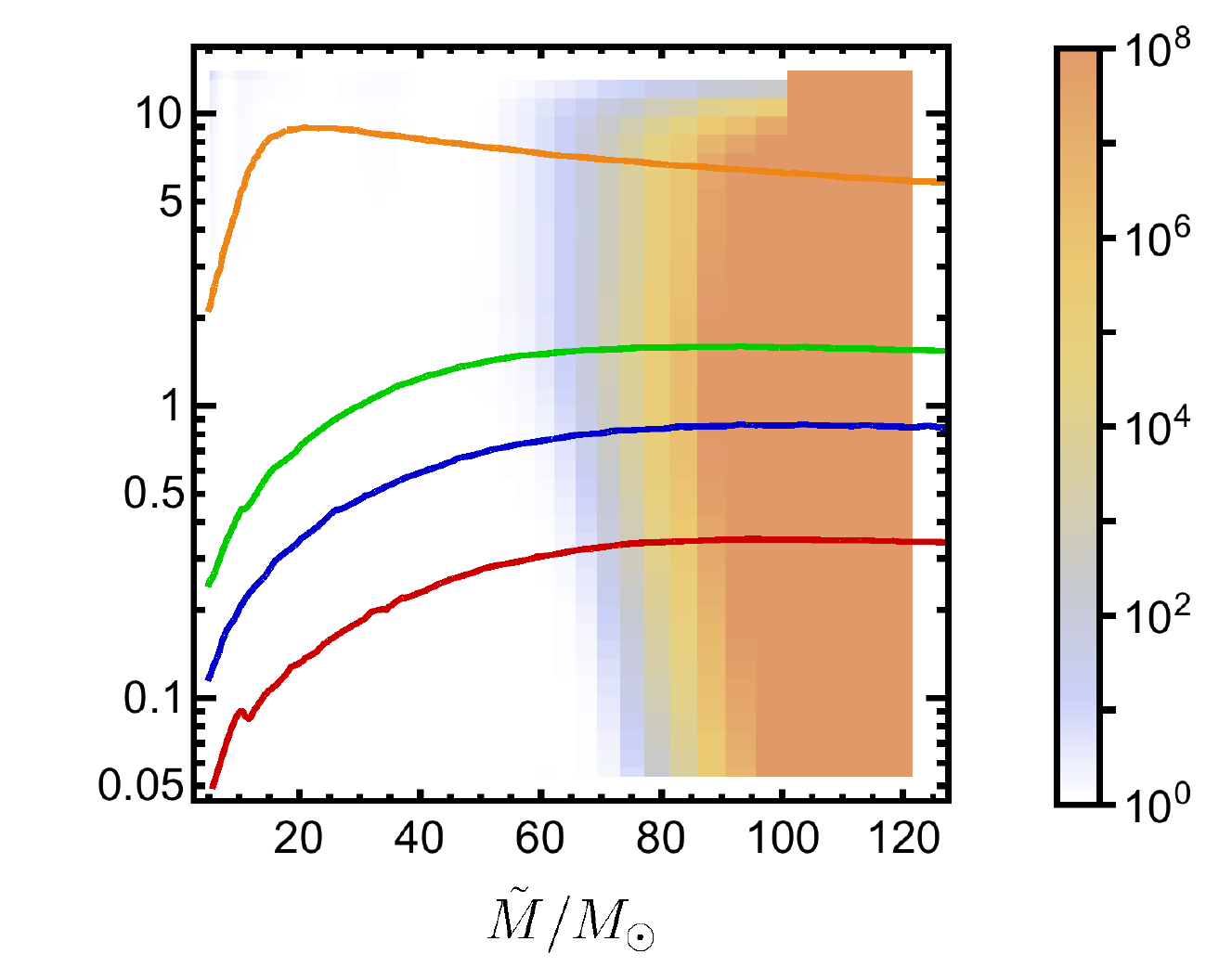}
\end{minipage}\\

\vspace{0.7cm}
\hspace{-0.2cm}
\begin{minipage}[]{0.48\linewidth}
\includegraphics[scale=0.43]{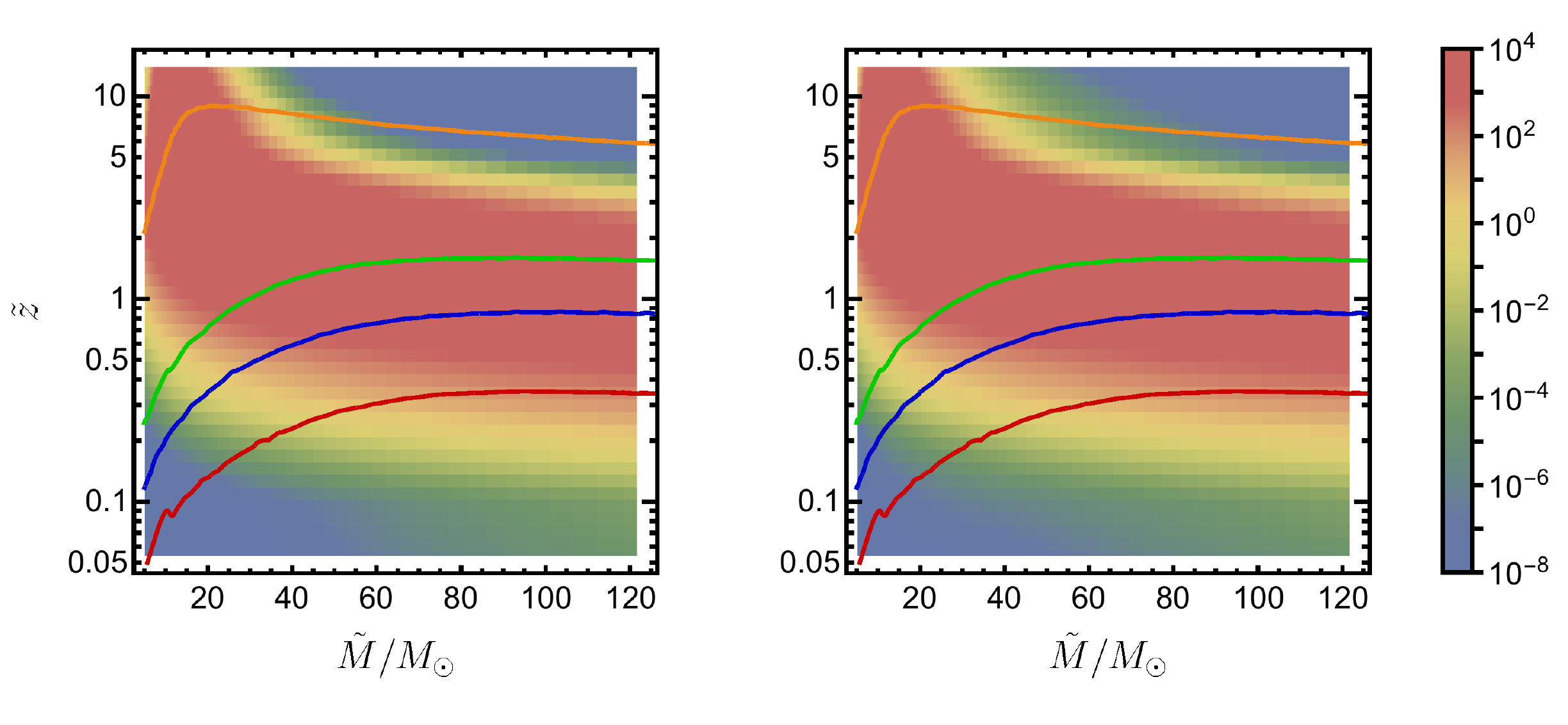}
\end{minipage}
\vspace{-1cm}
\hspace{0.6cm}
\begin{minipage}[]{0.48\linewidth}
\includegraphics[scale=0.43]{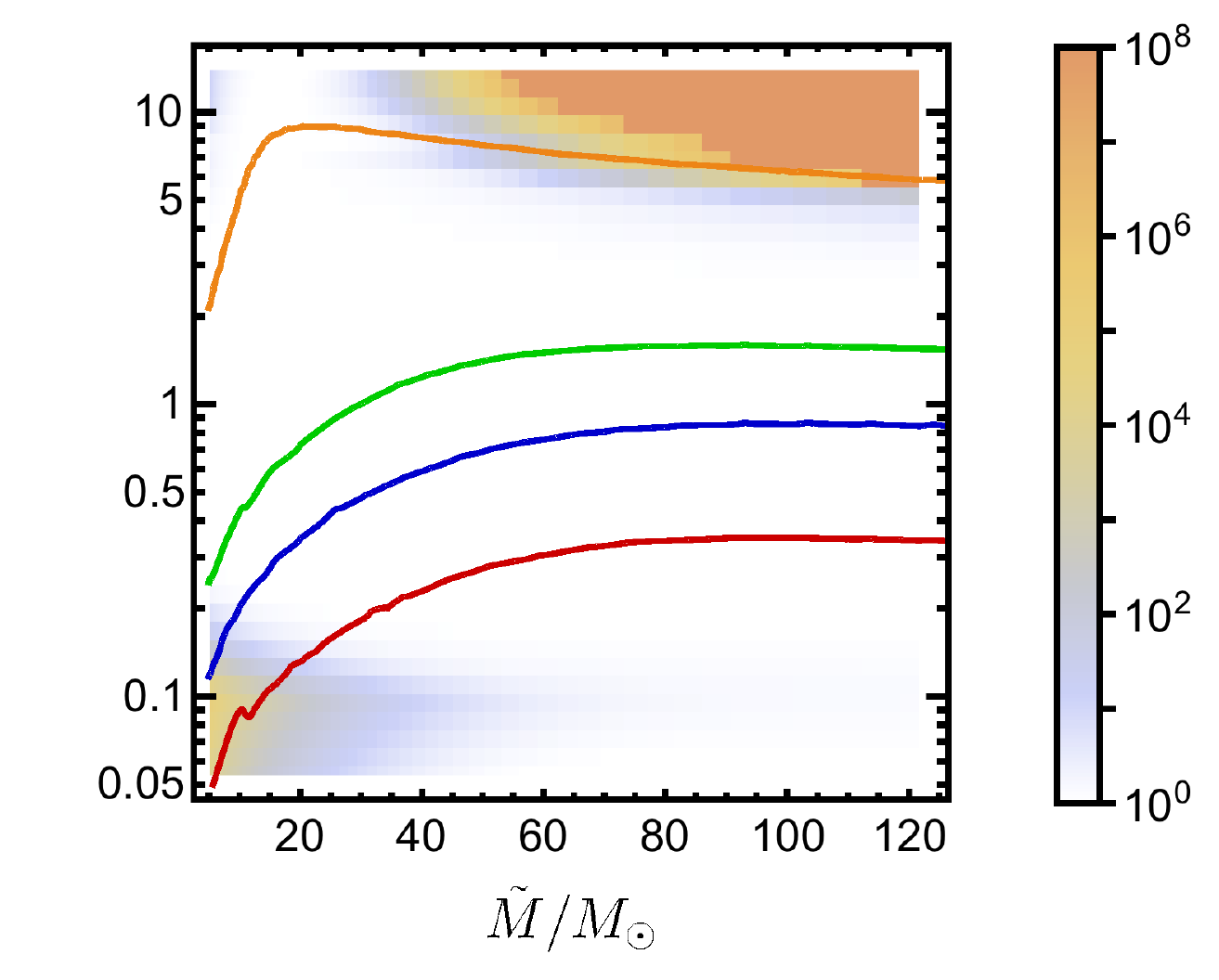}
\end{minipage}\\

\vspace{0.7cm}
\hspace{-0.2cm}
\begin{minipage}[]{0.48\linewidth}
\includegraphics[scale=0.43]{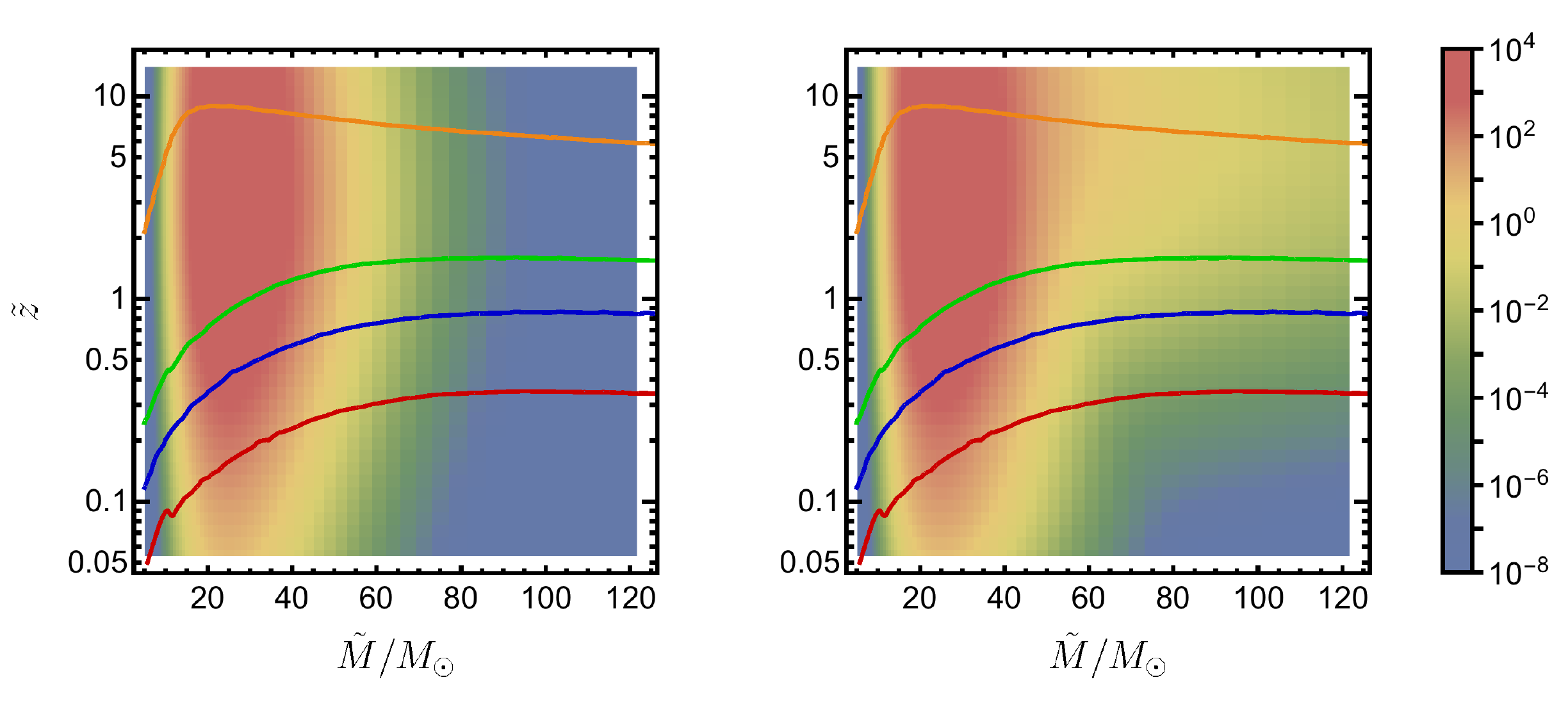}
\end{minipage}
\vspace{-1cm}
\hspace{0.6cm}
\begin{minipage}[]{0.48\linewidth}
\includegraphics[scale=0.43]{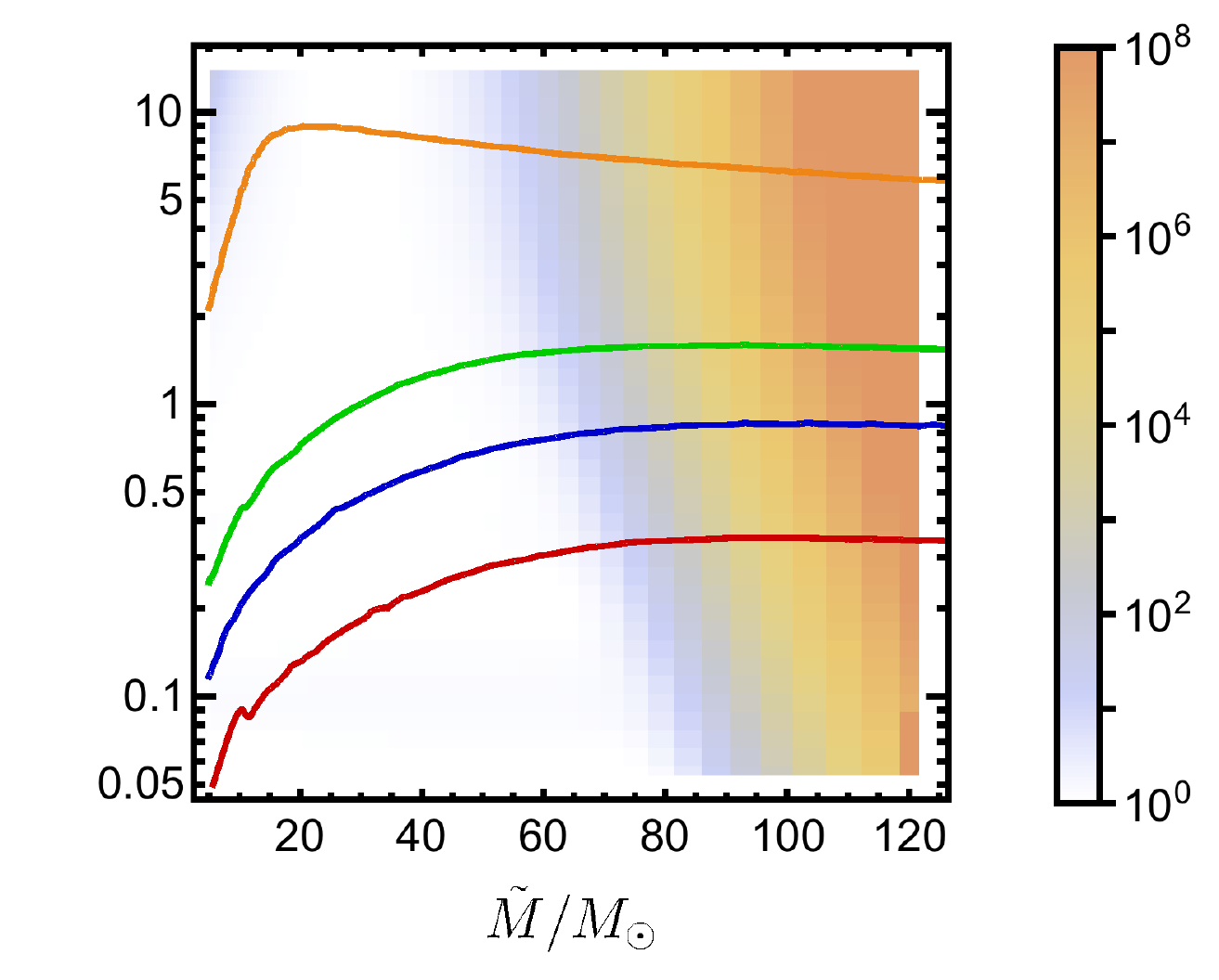}
\end{minipage}\\

\vspace{1cm}
\caption[]{Distribution of inferred mass $\tilde M$ and inferred redshift $\tilde z$ for BH-BH mergers $d^3N/(dt\,d\ln\tilde M\,d\ln\tilde z)\,[{\rm yr}^{-1}]$. From top to bottom we plot for models (1)-(4) in the order they are introduced in \reffig{zdist}. The case with lensing ({\it Middle}) is compared to the case without lensing ({\it Left}). Rate enhancement by lensing is also shown ({\it Right}). Each of the four detectors considered in \reffig{SNRcontour} cuts off the region according to the threshold curves over-plotted, with the same color coding adopted as before.}
\label{fig:obs_rate_schechter}
\end{figure*}

The top two rows of \reffig{detect_rate} show the distribution of intrinsic masses and redshifts of all detectable mergers for the two models. As in \reffig{detect_rate_pop3new}, all the detected mergers that lie beyond the threshold are strongly lensed. Figure \ref{fig:obs_rate_schechter} plots the distribution in the plane of {\em observed} parameters, along with the ratio between the rates in the cases with and without lensing for each bin (when interpreting \reffig{obs_rate_schechter}, note that each detector cuts off the plot at its individual threshold). We observe from the right-most plot of \reffig{obs_rate_schechter} that the lensed-fraction can be dominant for high BH masses $M \gtrsim 60\,M_\odot$. In our first model without PPSN/PSN, very massive systems exist at high redshifts, and magnification bias is only relevant at intermediate redshifts $0.1 < z <1$. In our second model, ultra-massive systems are forbidden by PPSN/PSN, and magnification bias is present for a wide range of redshifts. With the aid of lensing, LIGO is sensitive to a number of early mergers at $z\sim 2-5$, and can thus offer an insight into binary evolution in the early Universe. Since lensed and unlensed events are not distinguishable on an event-by-event basis without electromagnetic counterparts, one may only study the merger distribution with respect to the inferred parameters \refeq{obs-rate-lensing}, as shown in \reffig{obs_rate_schechter}.

\begin{figure*}[htbp]
\centering

\hspace{-1.5cm}
\includegraphics[scale=0.6]{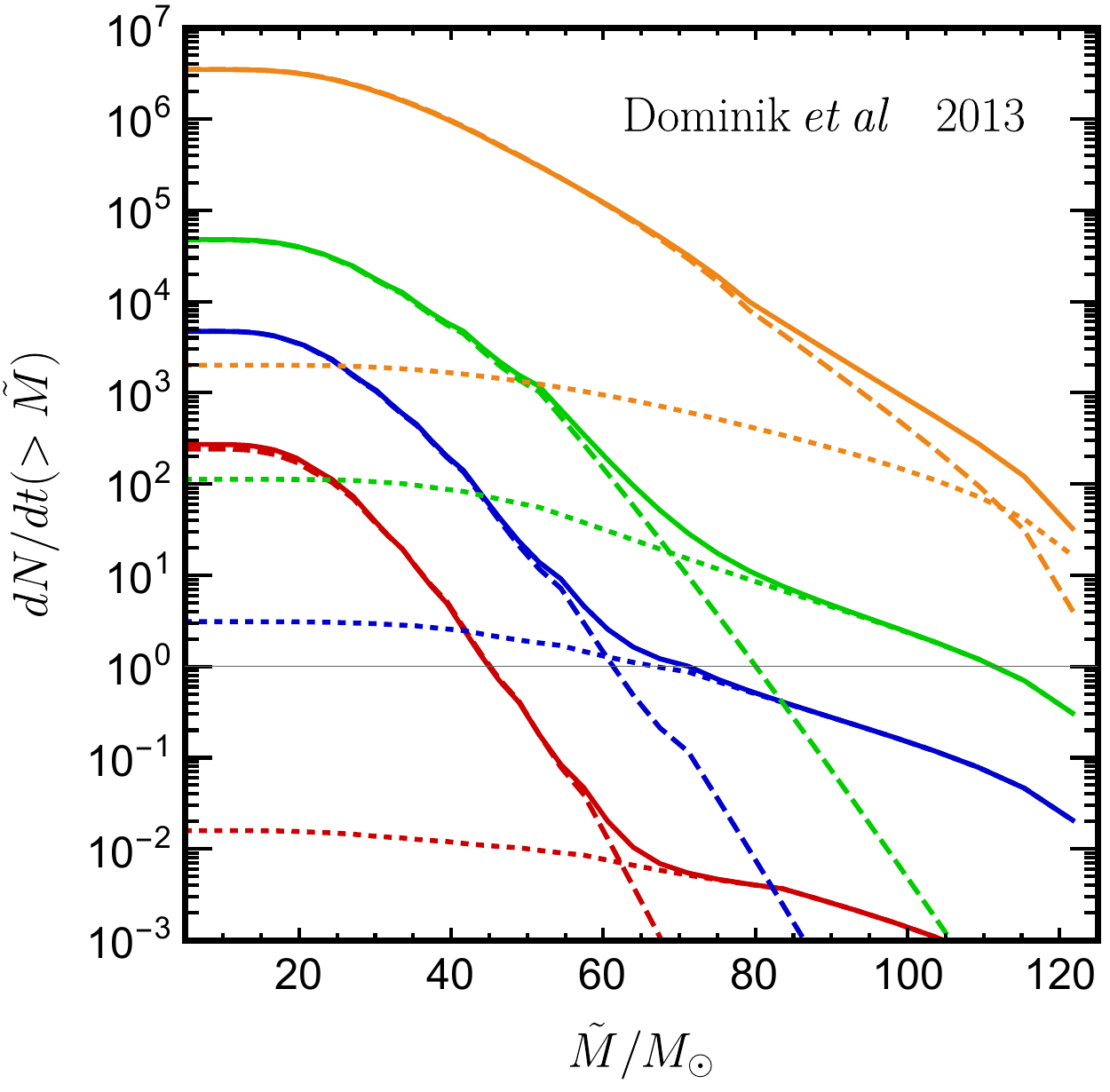}
\hspace{0.5cm}
\includegraphics[scale=0.6]{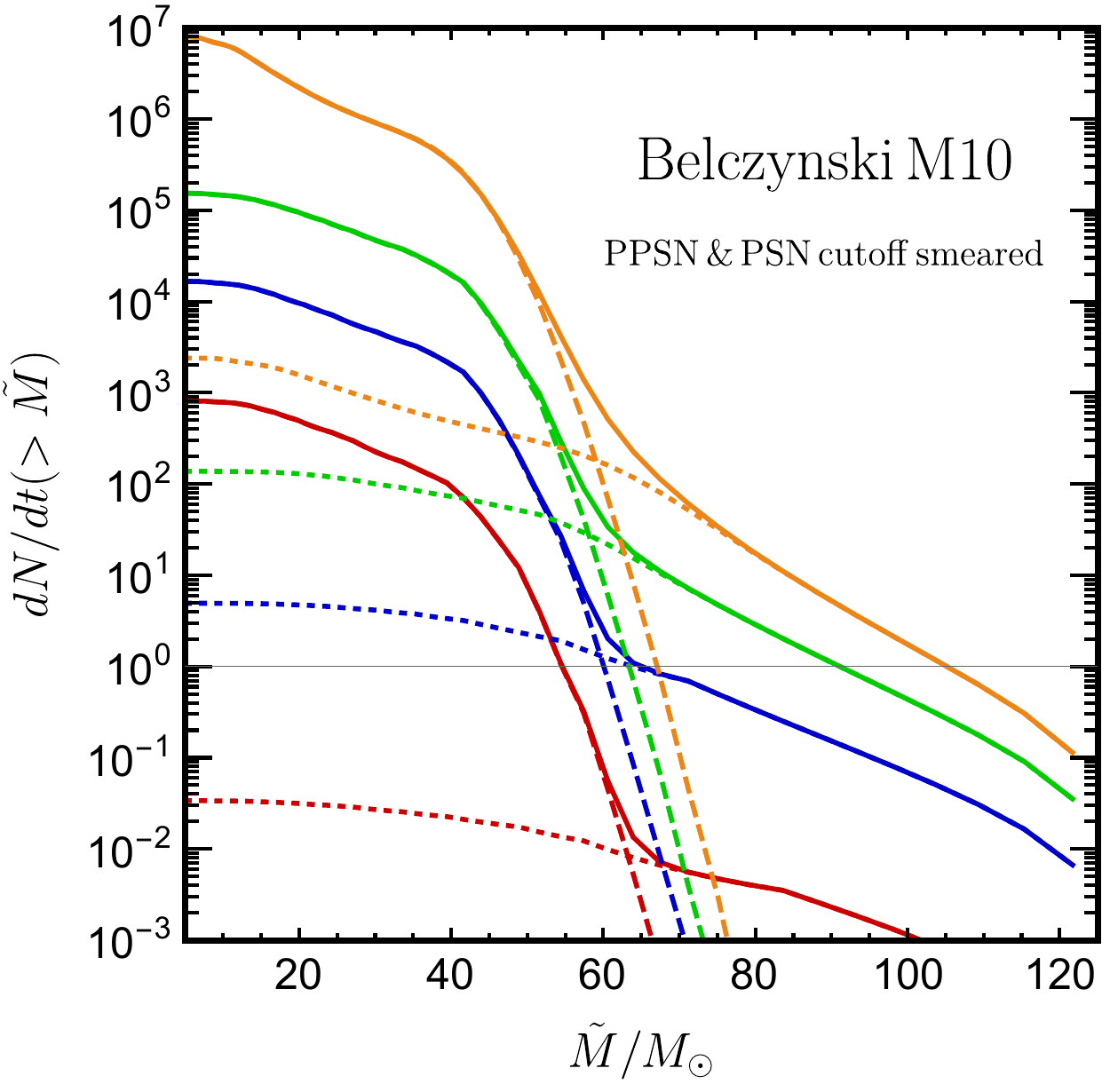}

\hspace{-1.5cm}
\includegraphics[scale=0.6]{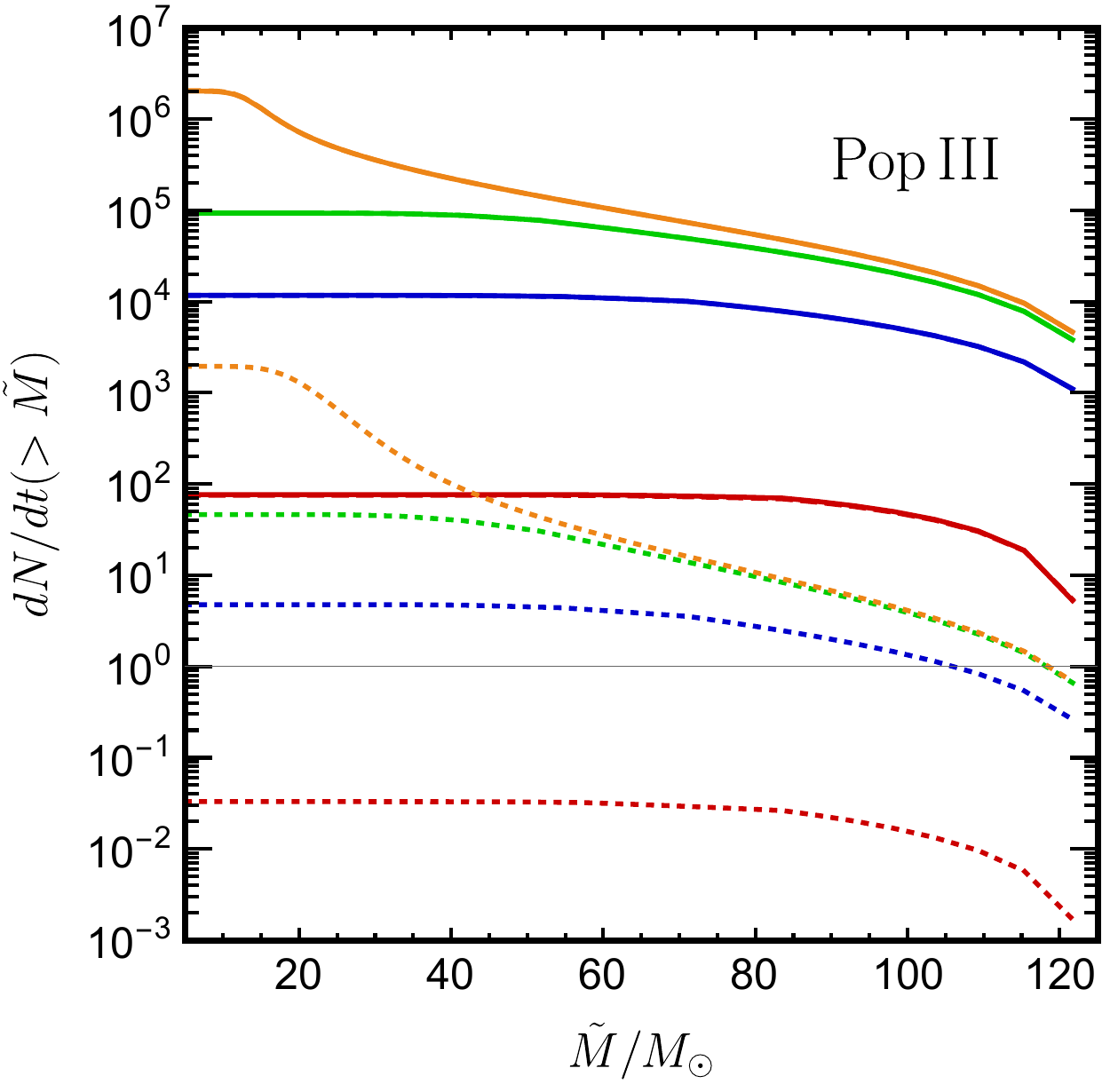}
\hspace{0.5cm}
\includegraphics[scale=0.6]{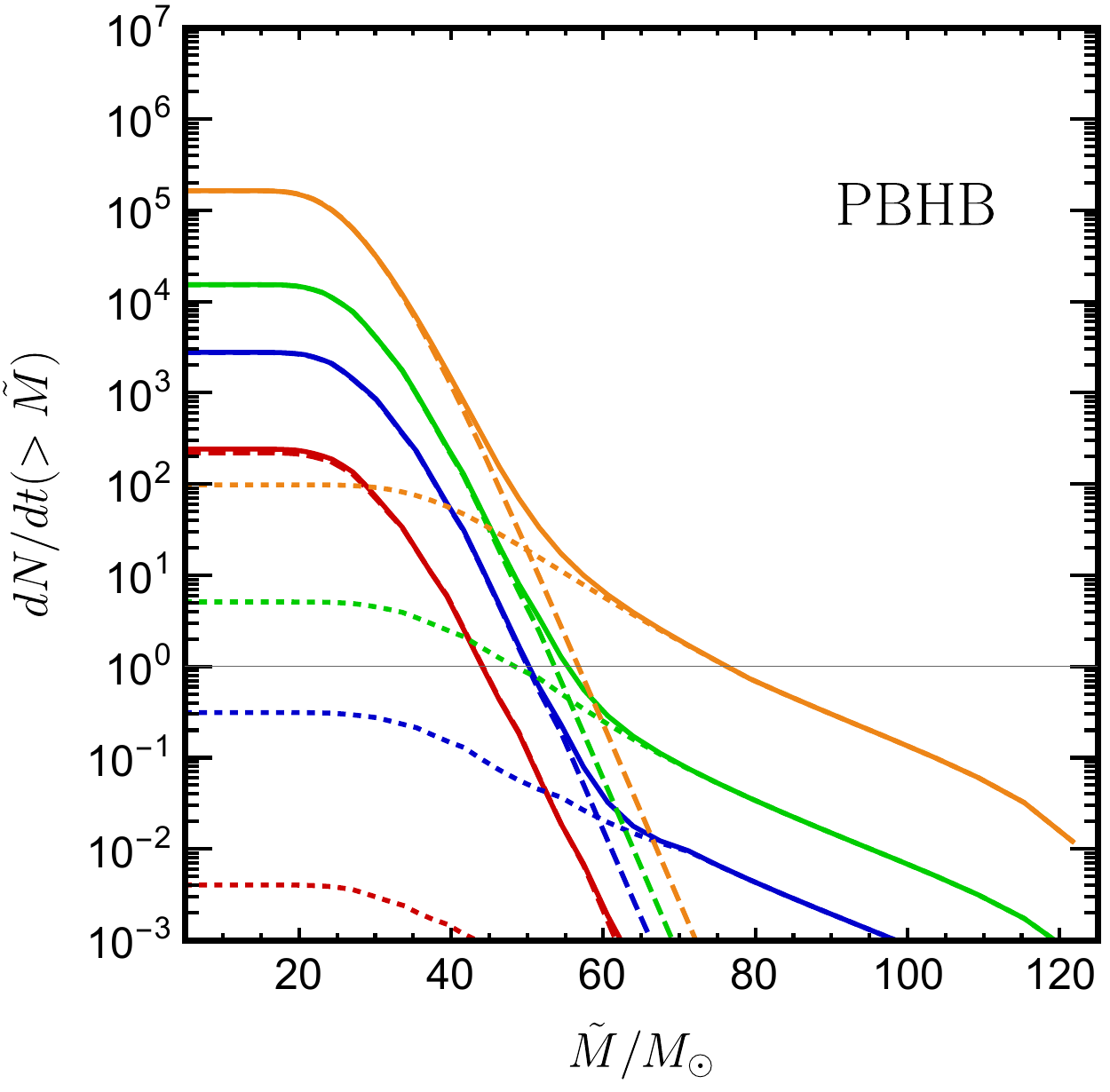}

\caption[]{Cumulative rate of detectable BH-BH mergers $dN/dt$ with mass scale larger than $M$ {\it per decade} of observation, for model (1)-(4) as introduced in \reffig{zdist}, respectively. The same color coding of \reffig{SNRcontour} is adopted for different detector sensitivities. We compare between the lensed distribution (solid), the unlensed distribution (dashed), and strongly lensed events with $\mu>3$ (dotted).}
\label{fig:obs_rate_schechter_Mdist}
\end{figure*}

Figure \ref{fig:obs_rate_schechter_Mdist} plots the projection of the distribution along the observed-mass axis. We see that lensing induces an ``ankle'' in the mass distribution at masses $M \simeq 60-80\,M_\odot$, which would be nontrivial to explain with a simple merger mass function without lensing. The effect is most pronounced for LIGO, which is luminosity limited, but may or may not be important for the ET. The lensing-induced tail's number count is too low to be detectable in a reasonable time-frame with {\tt current} LIGO's sensitivity, but are measurable after a few years of operation at {\tt design} or {\tt ultimate} sensitivities. 

\subsection{Mergers from Pop III stars}
\label{sec:pop3}

Pop III stars are a hypothesized generation of massive and short-lived stars that could have thrived in metal-free environments around $z \sim 10$~\cite{Bromm:2003vv}. Moreover, it has been proposed that Pop III stars of masses $25< M/M_\odot<140$ or $M/M_\odot > 260$ end their lives in massive BHs~\cite{Heger:2001cd}. The distribution of these massive BHs is open to speculation, but it is reasonable to expect that a significant fraction reside in binaries, either due to their parent stars' distribution~\cite{Stacy:2012iz}, or due to their evolution towards the center of their galactic potential under dynamical friction and subsequent capture~\cite{Madau:2001sc}. In this section, we consider the GW signature due to the hierarchical merger of these ancient binaries (a process that is reminiscent of our toy model in Sec.~\ref{sec:statistics}). Note that merger scenarios involving Pop III remnants have also been studied in Refs.~\cite{Kinugawa:2014zha, Hartwig:2016nde}.

The analysis in Ref.~\cite{Wyithe:2003ju} fixes the normalization $dn(z)/dt_s$ using a merger tree simulation that is tuned to produce the right mass in supermassive BHs, assuming that Pop III remnant BHs account for some fraction $f_{\rm MBH}$ of this mass (with the rest contributed by gas accretion). In the following, we assume $f_{\rm MBH}=0.01$, and infer the intrinsic merger rate density (see lower panel of \reffig{zdist}) from Ref.~\cite{Wyithe:2003ju} for stellar mass seeds (left panel of Figure 2 therein). We start with a log-normal distribution for initial seed masses centered at $15\,M_\odot$, with a width of $6\,M_\odot$ (as earlier, this corresponds to a BH mass budget $\Omega_{BH} \simeq 3\times 10^{-8}$). We fit the evolved mass distribution $dP(z)/dM$ to a log-normal distribution at each redshift.

Figure \ref{fig:detect_rate} shows that lensing magnification brings a significant number of mergers from $z>2$ (especially early-stage mergers between low-mass seeds) above the detection threshold for LIGO and its upgrades. However, these lensed mergers are buried under a population of detectable late-stage high-mass mergers from around $z \sim 1-2$ (as shown in \reffig{obs_rate_schechter}). The lensed events dominate only at high masses $\tilde M>40\,M_\odot$ at $\tilde z>5$, or at low masses $\tilde M<40\,M_\odot$ at $\tilde z <0.2$; the event rate is undetectably low in both ranges. We can also see this in the cumulative mass distribution in \reffig{obs_rate_schechter_Mdist}, which decays smoothly toward high masses $M \gtrsim 100\,M_\odot$ and shows no sign of magnification bias. This is due to a combination of the log-normal tail of high-mass mergers at given $z$ and an increase of individual BH mass on average as hierarchical merging progresses, which is to be contrasted with the Schechter cutoff and a suppression of recent high-mass mergers in \refsec{schechter}. In general, an extended heavy tail in the intrinsic mass distribution can wash out the strongly lensed contribution. Note also that the merger activity as predicted in Ref.~\cite{Wyithe:2003ju} finishes late ($z \sim 1$). Were this to end earlier, magnification bias would become more important.

\subsection{Mergers of primordial BHs}
\label{sec:primordial-BH-bin}

A relic population of black holes from the primordial Universe is a third possibility that has been considered for the origin of massive BH mergers~\cite{Bird:2016dcv,Clesse:2016vqa,Sasaki:2016jop}. We evaluate the GW signal from mergers within this model, but do not speculate about the origin of the relics. Proposals in the literature include direct collapse from horizon-scale peaks in the cosmological density field during radiation domination~\cite{Carr:1974bh}, or collapse due to the reduction in pressure support during cosmological phase transitions~\cite{Jedamzik:1996mr}. 

The model in Ref.~\cite{Bird:2016dcv} invokes the following assumptions: a) primordial BHs with masses $M \simeq O(10) \times M_{\odot}$ account for a substantial fraction of dark matter in the present Universe, and b) the BHs form close binaries in dense, low-velocity mini halo environments via dynamical capture, and quickly merge within one Hubble time. 

Since this process is most efficient in sub-galactic mass dark matter halos whose comoving abundance does not significantly evolve after $z \sim 10$ in the $\Lambda$CDM cosmology, we can assume a stationary merger rate density $dn(z)/dt_s$. Its exact value depends on the mass fraction of primordial BHs and the efficiency of dynamical capture. To facilitate comparison with other scenarios, we choose a value of $20~{\rm Gpc}^{-3}\,{\rm yr}^{-1}$. This is the merger rate visible to a detector that is limited to a local volume. 

Since only a small fraction of primordial BHs form binaries and merge, the merger mass distribution $dP(M)/dM$ may be assumed to be be stationary. Its specific form is open to speculation; if primordial BHs formed in a narrow redshift-range, we expect this to be a peaked distribution. In order to explain GW150914, we consider a log-normal distribution centered at $M_0=25\,M_\odot$,
\ba
\hspace{-0.5cm}
P_{\rm PBH}(M) & = & \frac{1}{\sqrt{2\pi}\,\sigma_{\ln M}\,M}\,\exp\left( - \frac{\left[ \ln(M/M_0) \right]^2}{2\sigma^2_{\ln M}}\right),
\ea
with a width parameter $\sigma_{\ln M} = 0.18$, corresponding to a small width of $\Delta M \sim 10\,M_\odot$. If the width were much larger, we would expect LIGO to have preferentially detected mergers more massive than $30\,M_\odot$.

Due to the relatively narrow BH mass-distribution, the lensing magnification significantly extends the redshift reach of LIGO and its upgrades (as can be seen in the lower row of \reffig{detect_rate}). Figure \ref{fig:obs_rate_schechter} shows that lensed events dominate the high-mass end of the inferred mass distribution over a wide range of inferred redshifts. If the mass distribution were much wider, the high magnification tail would be washed out. Note however that the total detectable number of high-mass events is low for even {\tt design} or {\tt ultimate} LIGO sensitivities, as can be seen in \reffig{obs_rate_schechter_Mdist}.  This is due to the assumption of a constant normalization for the rate with redshift (as noted earlier, the normalization is fixed to match the inferred rate from GW150914). In such a scenario, the ET can measure the lensing-induced tail and probe the intrinsic BH mass distribution.

\section{Discussion}
\label{sec:discussion}

In this paper we pointed out an intrinsic observational degeneracy between the masses and redshifts of binary BH mergers and their lensing magnifications in case that no electromagnetic counterparts, and hence redshifts, are available. Even though BH mergers are standard sirens, this prevents gravitational wave observers from determining lensing magnifications on an event-by-event basis. Traditional astrophysical sources from the high-redshift universe, such as galaxies, quasars and supernovae, do not suffer from this ambiguity because their electromagnetic spectra contain absolute scales that are set by non-gravitational physics.

If we use a standard cosmology for parameter estimation, the distribution of observed BH mergers in mass and redshift will differ from the intrinsic distribution. In particular, apparently high-mass mergers from low redshifts may be artifacts of lensing magnification. This effect is particularly important if BH mergers originate from standard stellar evolution, or from a low-mass primordial population.

Due to incompleteness originating from detector thresholds, it is not possible in general to deconvolve the effects of a given magnification PDF and recover the intrinsic event distribution. However, given a theoretical model of the merger mass distribution and its evolution, one can forward-model the effect of cosmic magnification by convolving with the appropriate PDF, and compare the resulting distribution to that of observed events. 

In this paper we have focused solely on the observed distribution, but in general we might be able to identify strongly lensed events through their multiplicity. Quite generically, multiple images exist in the regime of strong magnifications. The typically good angular resolution of astronomical surveys enables separation of the angular positions of multiple images on the sky. This is unlikely to be an option for GW observations, whose angular localization is too coarse to achieve the arcsecond precision needed for resolving the multiple images due to galaxy lenses. Due to the burst nature of the sources and the excellent temporal resolution of the observations, it is more likely that we can identify multiple images of mergers through their mutual time delay. Since multiple images produced by galaxy lenses are typically separated by time delays of weeks to months, those that are sufficiently bright will be detected as separate GW chirps. We expect that we can identify (with high statistical significance) events whose arrival directions and reconstructed dimensionless parameters are consistent, but whose apparent mass scales and redshifts are related by a consistent magnification ratio. Even in this case, we cannot uniquely pin down the absolute source redshift and mass scale due to the unknown absolute magnification scale. Nevertheless, a sample of multiply-imaged mergers identified in this way could constrain the properties of strong gravitational lenses, and conceivably offer a path toward estimating and correcting for the population of lensed black hole mergers.  We will present a detailed study of some of these prospects in a forthcoming publication.
 
\begin{acknowledgments}

We thank John Miller for sharing the noise curve for a possible ultimate upgrade to LIGO, Tobias Baldauf for insightful discussions about lensing statistics, and  Neil Cornish and Aaron Zimmerman for very helpful comments on an earlier version of this manuscript. We are especially thankful to Krzysztof Belczynski for sharing with us the results of the latest population synthesis simulation performed with the {\tt StarTrack} code.

LD is supported at the Institute for Advanced Study by NASA through Einstein Postdoctoral Fellowship grant number PF5-160135 awarded by the Chandra X-ray Center, which is operated by the Smithsonian Astrophysical Observatory for NASA under contract NAS8-03060. TV acknowledges support from the Schmidt Fellowship and the Fund for Memberships in Natural Sciences at the Institute for Advanced Study. KS gratefully acknowledges support from the Friends of the Institute for Advanced Study. The research of KS is supported in part by a Natural Sciences and Engineering Research Council of Canada (NSERC) Discovery Grant.

\end{acknowledgments}

\appendix

\section{Wave effects at high magnification}
\label{app:waveeffects}

In this appendix, we consider the question of whether geometrical, or ray optics describes gravitational lensing of GWs by galaxy lenses. The approximation breaks down when distinct rays intersect, which naturally occurs when sources approach caustics on the source-plane \citep{1983ApJ...271..551O}. Large magnifications $\mu$ generally occur in this region of parameter space. Since the results in the body of the paper depend on the cross-section for strong-lensing or large $\mu$, we need to verify whether geometrical optics holds over this domain. 

Previous studies of lensing of GWs from supermassive BH binaries have shown that wave-effects kick in at $\mu \gtrsim \mathcal{O}(10)$ where there is significant weight in the magnification PDF, and hence lead to observable effects in the waveforms (see Refs.~\cite{Takahashi:2003ix, Sereno:2011ty, 2014PhRvD..90f2003C}). In this appendix, we will show that GWs from stellar mass BH binaries are typically at higher frequencies and thus are lensed geometrically over the domain of the magnification PDF.

When the geometrical approximation breaks down, wave effects both regulate the magnification, and leave characteristic imprints in the observed waveforms. We will see that the phenomenology of wave-effects is much simpler for GWs from stellar-mass BH binaries -- this is because the geometrical optics validity of the the Since our primary aim here is to check the validity of our magnification PDFs for stellar-mass BH binaries, and not to explore the diverse phenomenology of wave-effects, we adopt the following procedure.

The usual treatment starts by deriving the diffraction integral for a scalar-valued wave traversing multiple lens planes under the paraxial approximation \citep{2003astro.ph..9696Y}. The geometrical-optics approximation reduces the diffraction integral to the lens equation with multiple deflections. Since our primary aim here is to check the validity of our magnification PDFs for stellar-mass BH binaries, we adopt the following simplifying strategy: (a) we know that large magnifications ($\mu \gtrsim 10$) typically occur when the source is close to the fold caustic of a single lens plane (as can be seen from the asymptotic power-law for the PDF $dP/d\mu \sim \mu^{-3}$ in Figures \ref{fig:muPDF} and \ref{fig:muPDFtail}). (b) we check for wave effects in such single-plane lensing, and show that for GWs from stellar-mass BBHs, geometrical optics holds until $\mu \gtrsim \mu_{\rm max} \gg 10$ (if $\mu_{\rm max} \lesssim 10$, we have to consider multi-plane lensing). (c) We then show that the cumulative probability $P(\mu \gtrsim \mu_{\rm max}) \ll 1$.

We follow the derivation in Ref.~\cite{1992grle.book.....S} for wave-effects in single-plane lensing. Under the thin-lens approximation, the observed amplitude obeys a Fraunhofer diffraction equation, i.e., it is an integral over the entire lens plane with each point, $\bt x$, weighted by a factor $\exp{[i f \phi(\bt x)]}$. Here, $\phi(\bt x)$ is the dimensionless Fermat-potential (scaled time-delay), and $f$ is the dimensionless numerical factor
\begin{align}
  f & = \frac{2 \pi \nu}{c} \frac{\chi(z_{\rm s}) \chi(z_{\rm d})}{[ \chi(z_{\rm s}) - \chi(z_{\rm d}) ]} \left( \frac{\xi_0}{D_{\rm d}} \right)^2 \mbox{,} \label{eq:fraunhoferf}
\end{align}
where $\nu$ is the observed frequency, $\xi_0$ is a physical length-scale on the lens plane, $z_{\rm s}$ and $z_{\rm d}$ are the source and lens redshifts, respectively, and $\chi(z)$ is the comoving distance to redshift $z$. 

For a numerical estimate, we choose $\xi_0$ to be the Einstein radius for a singular isothermal lens with velocity dispersion $\sigma_v^2$, which equals $4\pi (\sigma_v^2/c^2) (\chi(z_{\rm s}) - \chi(z_{\rm d}))/\chi(z_{\rm s})$ (this is only for obtaining a numerical estimate; the lenses responsible for the results in Sec.~\ref{sec:mpdf} are not singular isothermal lenses).  Substitution into the above equation yields
\begin{align}
  f & = \frac{32 \pi^3 \nu}{c} \left( \frac{ \sigma_v}{c} \right)^4 \frac{ \chi(z_{\rm d}) }{ \chi(z_{\rm s}) } \left[ 1 - \frac{ \chi(z_{\rm d}) }{ \chi(z_{\rm s}) } \right] \chi(z_{\rm s}) \nonumber \\
  & = 1.11 \times 10^9 \times \left( \frac{ \chi(z_{\rm d})/\chi(z_{\rm s}) \left[ 1 - \chi(z_{\rm d})/\chi(z_{\rm s}) \right] }{ 0.25 } \right) \times \nonumber \\
  & ~~~\left( \frac{\nu}{100 \ {\rm Hz} } \right) \left( \frac{ \sigma_v}{161 \ {\rm km \ s^{-1}} } \right)^4 \left( \frac{\chi(z_{\rm s})}{\chi(z_{\rm s}=2) = 5.27 \ {\rm Gpc}} \right) \mbox{.} \label{eq:ffactor}
\end{align}
The factor $f$ is numerically large; typically, we can expand the time-delay as a quadratic function around normal image locations and evaluate a Gaussian integral to get the observed amplitudes. When the source is at a caustic, the quadratic term vanishes and the magnification is formally infinite. If the caustic is a fold, the time-delay behaves as a cubic function of image location, and the effective magnification is corrected to an Airy function along the lensing map's trivial direction (see \S 7.3 of Ref.~\cite{1992grle.book.....S}). The maximum magnification is
\begin{align}
  \mu_{\rm max} 
  & = 4 \pi f^{1/3} \biggr\vert \frac{ (1 - \kappa) }{ {\bt T} \cdot A^{\rm T} A \cdot {\bt T}} \biggr\vert^{1/3} Q^2 \mbox{,} \ Q \approx 0.5357 \mbox{,} \label{eq:magcutofffold}
\end{align}
where $\kappa$, $A$, and $\bt T$ are the dimensionless surface-mass density, Jacobian of the lens map, and tangent to the critical curve, respectively. Substituting in the value of $f$ from Eq.~\eqref{eq:ffactor}, we obtain
\begin{align}
~~~~ & \!\!\!
  \mu_{\rm max} \nonumber \\
  & = 3.74 \times 10^3 \times \biggr\vert \frac{ (1 - \kappa) }{ {\bt T} \cdot A^{\rm T} A \cdot {\bt T} } \biggr\vert^{1/3} \times \nonumber \\
  & ~~~ \left( \frac{ \chi(z_{\rm d})/\chi(z_{\rm s}) \left[ 1 - \chi(z_{\rm d})/\chi(z_{\rm s}) \right] }{ 0.25 } \right)^{1/3}\left( \frac{\nu}{100 \ {\rm Hz} } \right)^{1/3} \times \nonumber \\
  & ~~~\left( \frac{\sigma_v}{161 \ {\rm km \ s^{-1}} } \right)^{4/3} \left( \frac{\chi(z_{\rm s})}{5.27 \ {\rm Gpc}} \right)^{1/3} \mbox{.}
  \label{eq:mumaxgeom}
\end{align}
The dimensionless factors are order unity, except near cusps, which generically form a small (i.e., measure-zero) subset of the caustics. The magnifications quoted in the body of the paper are well below this limit. In fact, none of the results from the numerical simulations have statistics in this regime. If we use our fitting formula for $dP/d\ln\mu$ (Eq.~\eqref{eq:semi-analytical-dai}), events above $\mu_{\rm max}$ have a cumulative probability of $4 \times 10^{-10}$ for the highest source-redshift we considered ($z_{\rm s} = 20$). Thus, essentially all the observable strong-lensing events that contribute to the degeneracy in the main text are described by the geometrical approximation\footnote{Note that point-source magnifications are regulated by an integration over the brightness profile for extended sources (we treat binary black holes as point sources of GWs).}.

We can also easily check that this conclusion is insensitive to details of the source and lens modeling. From Eq.~\eqref{eq:mumaxgeom}, GWs with $\nu \sim 10^{-6} {\rm Hz}$ have a threshold $\mu_{\rm max}$ of $O(10)$ (which would result in a measurable rate of lensed events with wave effects). These GWs would be out of the LIGO band, and hence would not be detectable using ground-based interferometers. The use of the singular isothermal lens model for numerical estimates is not crucial too -- assuming a different $\kappa$ model for galaxy lenses will not bias the Einstein radius $\xi_0$ in Eq.~\eqref{eq:fraunhoferf} by more than an order of magnitude, and the maximum magnification in Eq.~\eqref{eq:magcutofffold} depends only sub-linearly on $\xi_0$.

An exception to the discussion above is microlensing by stars, where the Einstein radius $\xi_0$ is small compared to that of galaxy or cluster lenses (and hence $1 \lesssim \mu_{\rm max} \ll \mu_{\rm max, gal} \simeq 3.7 \times10^3$). This can lead to interesting wave-effects in the lensing of GWs from stellar-mass BBHs (see Refs.~\cite{2004A&A...422..761M, 2007A&A...463...31M}), but these effects do not change the degeneracy that is the subject of this paper.

One last caveat is that the scalar-wave treatment above neglects non-trivial couplings to background space-time curvature that are specific to spin-two metric waves, i.e., GWs \citep{blandford2003applications}. By the equivalence principle, curvature coupling describes wavelength-dependent corrections to gravitational-wave propagation of $\mathcal{O}([c/(2\,\pi\,\nu\,R_c)^2]\,h_B)$, where $\nu$ is the wave frequency, $h_B$ is the amplitude of the background metric fluctuation, and $R_c$ is the spatial scale over which the background metric varies. As we have seen above, even in the {\it absence} of the curvature coupling, diffraction corrects ray optics at $\mathcal{O}(1/f)$, where $f$ is the dimensionless constant in \refeq{fraunhoferf}. The constant $f$ can be written as $f \sim (2\,\pi\,\nu/c)\,R_g$, where $R_g$ is the deflector's typical gravitational radius, which is smaller than but at most equal to $R_c$. Therefore, diffractive corrections dominate over the effects of curvature coupling, and since the former are negligible for the GWs of interest, we are justified in neglecting the latter.

\section{Fit for magnification PDF}
\label{app:lenPDFfitLiang}

We propose a fit for the source magnfication PDF in the form of a log-normal distribution convolved with a heavy-tailed kernel
\ba
\label{eq:semi-analytical-dai}
\frac{dP(\mu)}{d\ln\mu} & = & F(\mu; t_0,\lambda,\delta) = A(t_0)\,\int^{+\infty}_0\,dt\,\exp\left[\frac{\lambda}{t+t_0} - 2 t \right] \en
&& \times \frac{1}{\sqrt{2\pi}\,\sigma}\,\exp\left[ -\frac{\left(\ln\mu - \delta - t \right)^2}{2\sigma^2} \right].
\ea
We shall focus on the special choice $\lambda=5$, which we observe provides a good fit of the realistic magnification PDF. Apart from the parameter $\lambda$, the function has a couple of free parameters: $\sigma$ characterizes the width of the log-normal distribution, $\delta$ is a shift parameter, and $t_0$ controls the relative size of the heavy tail. We demand that $dP/d\ln\mu \equiv F(\mu;t_0,\lambda,\delta)$ is a properly normalized probability distribution for the magnification factor $\mu$, i.e.  $\int\,d\ln\mu\,F(\mu;t_0,\lambda,\delta)=1$. This fixes the normalization factor $A(t_0)$ to

\begin{table}[t]
\begin{center}
\setlength\tabcolsep{9pt}
\begin{tabular}{l|c|c|c}
\specialrule{.1em}{.05em}{.05em} 
 & $\sigma$ & $e^{-\delta}$ & $t_0$ \\
\hline
\hline
$z=0.7$ & 0.008 & 1.0380 & $ 0.365 $ \\
$z=1$ & 0.010 & 1.0465 & $ 0.399 $ \\
$z=2$ & 0.028 & 1.0700 & $ 0.471 $ \\
$z=3$ & 0.050 & 1.0859 & $ 0.511 $ \\
$z=5$ & 0.078 & 1.1065 & $ 0.557 $ \\
$z=10$ & 0.110 & 1.1327 & $ 0.609 $ \\
$z=20$ & 0.150 & 1.1649 & $ 0.666 $ \\
\specialrule{.1em}{.05em}{.05em} 
\end{tabular}
\caption{\label{tab:muPDF_par_Dai} Parameters for \refeq{semi-analytical-dai} at a number of representative source redshifts. The value for $t_0$ is obtained by matching with the strong lensing optical depth $\tau(\mu>10)$ found in Ref.~\cite{Hilbert:2007jd}.}
\end{center}
\end{table}

\ba
A(t_0) = \left[ \int^{+\infty}_0\,dt\, \exp\left[\frac{\lambda}{t+t_0} - 2 t \right] \right]^{-1}.
\ea 
The further requirement of unit mean magnification $\VEV{\mu} = \int\,d\ln\mu\,\mu\,F(\mu;t_0,\lambda,\delta)=1$ uniquely fixes the shift parameter $\delta$ in terms of $t_0$ and $\sigma$.
\ba
e^{-\delta} = A(t_0) \int^{+\infty}_0\,dt\, \exp\left[\frac{\lambda}{t+t_0} - 2 t \right]\,e^{t+\frac{\sigma^2}{2}}.
\ea
We now show that the semi-analytical form \refeq{semi-analytical-dai} reproduces the correct high-magnification tail $dP/d\ln\mu \propto \mu^{-2}$. In the limit of large $\mu$, the log-normal function can be replaced by a narrow Dirac-delta function
\ba
\frac{1}{\sqrt{2\pi}\,\sigma}\,e^{ -\left(\ln\mu - \delta - t \right)^2/(2\sigma^2)} \rightarrow \delta_D\left(t - \ln\mu - \delta\right),
\ea
so that the $t$-integral becomes
\ba
\label{eq:tailapprox}
\hspace{-0.5cm}
F(\mu; t_0,\lambda,\delta) & \rightarrow & A(t_0)\,\mu^{-2}  \,\exp\left[\frac{\lambda}{\ln\mu+\delta+t_0}-2\delta\right] \en
& \approx & A(t_0)\,\mu^{-2}  \,e^{\lambda/\ln\mu}.
\ea

\begin{figure}[t]
\centering
\hspace{-0.5cm}
\includegraphics[scale=0.65]{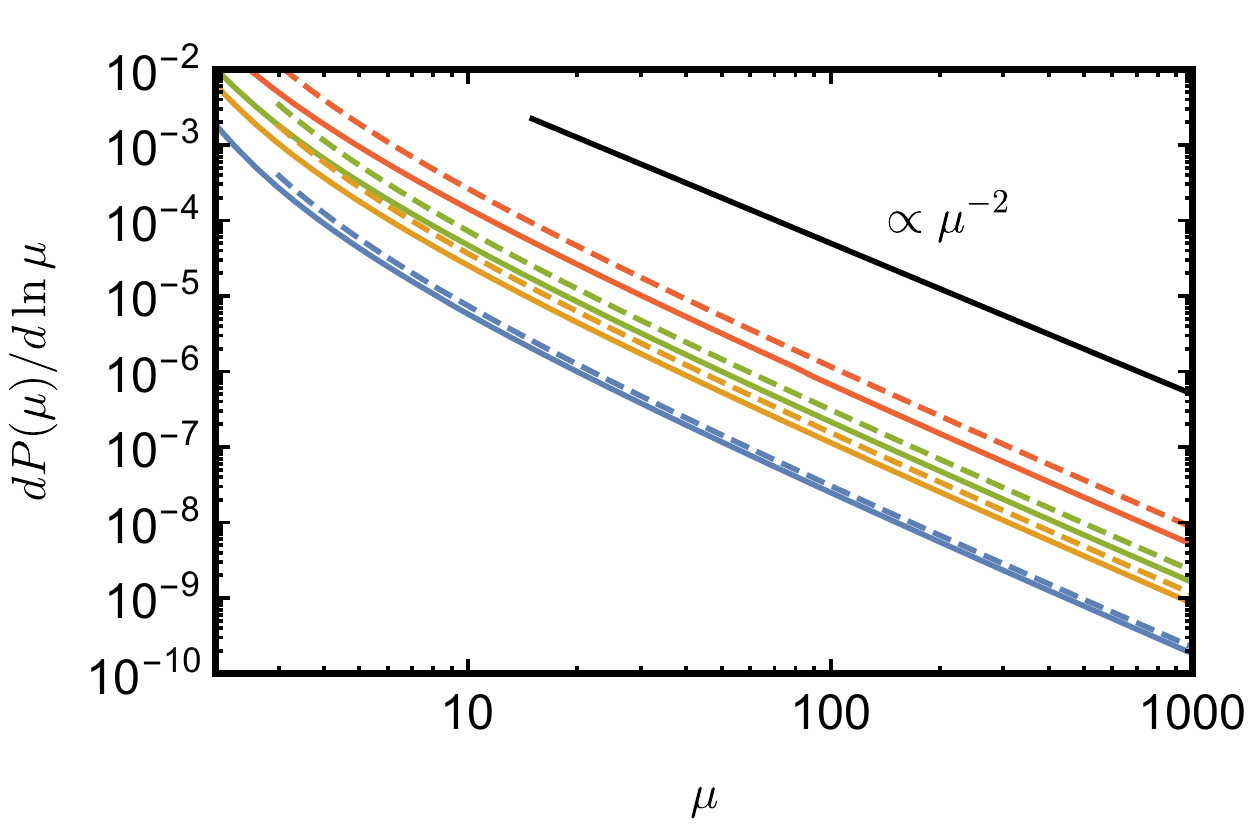}
\caption[]{Large-$\mu$ tail of parametric fit \refeq{semi-analytical-dai}. From bottom to top we plot for representative source redshifts $z=1,\,2,\,3,\,10$. Leading approximation for asymptotic behavior using \refeq{tailapprox} is shown in dashed curves. Higher order corrections account for remaining discrepancies.}
\label{fig:muPDFtail}
\end{figure}

Numerically, $|\delta|\ll 1$ and $t_0 \sim \mathcal{O}(1)$. For sufficiently large $\mu$, the $\ln\mu$-dependence is unimportant, and the distribution asymptotes to a power law $\sim\mu^{-2}$. See \reffig{muPDFtail} for a comparison with the $\mu^{-2}$ law. Since $\delta$ and $t_0$ drop out of the asymptotic form as long as they are small, $t_0$ can be uniquely matched to a strong lensing optical depth $\tau(\mu>\mu_0)$ through
\ba
A(t_0) & = & c_0/\tau(\mu>\mu_0),
\ea
where we define the constant $c_0$, which depends on $\mu_0$ as,
\ba
c_0 & = & \int^{+\infty}_{\mu_0}\,\frac{d\mu}{\mu}\,\exp\left[ \frac{\lambda}{\ln\mu} - 2 \ln\mu \right].
\ea
We choose the threshold magnification to be $\mu_0=10$ in order to match the numerical optical depth of Ref.~\cite{Hilbert:2007jd}. This determines $t_0$ as a function of the source redshift $z$. We also obtain a smooth fit for $\sigma$ and $\delta$ as a function of $z$ by matching the weak lensing portion to the result of Ref.~\cite{Takahashi:2011qd}. In \reftab{muPDF_par_Dai}, we list the numerical values for these parameters for a number of source redshifts.

\bibliographystyle{apsrev4-1-etal}
\bibliography{references_teja,references_liang,references_kris}

\end{document}